\documentclass[]{mn2e}
\usepackage{graphicx}

\newcommand{\Tef}{T$_{\rm eff}$~}
\newcommand{\HHO}{H$_2$O~}

\newcommand{\CC}{$^{12}$C$^{16}$O~}
\newcommand{\CCC}{$^{13}$C$^{16}$O~}

\begin{document}

\title{Carbon Monoxide in low-mass dwarf stars}
\author[H. Jones et al.]{
\parbox[t]{\textwidth}{Hugh R.~A.~Jones$^{1,2}$, Yakiv Pavlenko$^3$,
Serena Viti$^4$, R.J. Barber$^4$, Larisa A. Yakovina$^3$,
David Pinfield$^1$, Jonathan Tennyson$^4$\\}
\vspace*{6pt} \\
$^1$ Centre for Astrophysics Research, University of Hertfordshire, 
College Lane, Hatfield, Hertfordshire AL10 9AB\\
$^2$ Astrophysics Research Institute, Liverpool John Moores University,
Twelve Quays House, Egerton Wharf, Birkenhead CH41 1LD\\
$^3$Main Astronomical Observatory of Academy of Sciences o
Ukraine, Golosiiv woods, Kyiv-127, Ukraine 03680\\
$^4$ Department of Physics and Astronomy, University College London,
Gower Street, London WC1E 6BT}

\date{Accepted ................. Received .............}

\maketitle

\begin{abstract}
We compare high resolution infrared observations of the CO 3-1 bands in the 
2.297--2.310 $\mu$m region of M  dwarfs and one L dwarf with theoretical 
expectations. We find a good match between the observational and synthetic 
spectra throughout the 2000-3500K temperature regime investigated. Nonetheless, 
for the 2500-3500 K temperature range the temperatures that we derive from 
synthetic spectral fits are higher than expected from more empirical methods 
by several hundred K. In order to reconcile our findings with the empirical 
temperature scale it is necessary to invoke warming of the model atmosphere 
used to construct the synthetic spectra. We consider that the most likely 
reason for the back-warming is missing high temperature opacity due to water
vapour. We compare the water vapour 
opacity of the Partridge \& Schwenke (1997) line list used for the model 
atmosphere with the output from a preliminary calculation by Barber \& Tennyson 
(2004). While the Partridge \& Schwenke line list is a reasonable spectroscopic 
match for the new line list at 2000~K, by 4000~K it is missing around 25\% of 
the water vapour opacity. We thus consider that the offset between empirical 
and synthetic temperature scales is explained by the lack of hot water vapour 
used for computation of the synthetic spectra. For our coolest objects with 
temperatures below 2500~K we find best fits when using synthetic spectra which 
include dust emission. Our spectra also allow us to constrain the rotational 
velocities of our sources, and these velocities are consistent with the broad 
trend of rotational velocities increasing from M to L.

\end{abstract}


\section{Introduction}
Low mass dwarf stars dominate our Galaxy in terms of
number. They provide a probe of our understanding of main
sequence stellar evolution and are the key in determining the boundary
between stellar and sub-stellar objects.  There are relatively few
observations of known-mass low mass stars.  Parameters such as
effective temperature and metallicity, vital in determining
positions in H-R diagrams, remain controversial.  To reliably
constrain the low-mass initial stellar mass function it is essential
to know the basic properties of standard low-mass M, L and T dwarfs.
A correct determination of the mass function relies on an accurate
transformation from luminosity and temperature to mass.  These
relationships are sensitive to the stellar chemical composition.
For hotter objects colour-colour diagrams are reasonably reliable  indicators of 
temperature and metallicity. However such diagrams for low mass
dwarfs do not yet reproduce the
broadband fluxes within a reasonable error and therefore cannot be
uniquely used to determine reliable temperatures, metallicities and gravities.
Ideally it would be useful to have spectroscopic signatures sensitive 
to temperature, metallicity and gravity that are reproducible with synthetic 
spectra. 

\par
Many authors have determined the properties of low-mass objects using
synthetic spectra. However, the use of 
such synthetic spectra are problematic because (1) the objects are
dominated by various diatomic and triatomic molecules whose high 
temperature properties
are poorly understood and (2) the large number of different
transitions means that most transitions are substantially blended with
other competing opacities. One potential route to resolve these issues
is to try to find spectral regions where these issues are less 
problematic. For example the middle of the J-band window is a promising
region (Jones et al. 1996; McLean et al. 2003). Although this 
region is relatively transparent and is in a wavelength regime where
infrared spectrometers are relatively sensitive it does have shortcomings.
In addition to the problems with modelling water vapour at 
short wavelengths (Jones et al. 2002), it is now clear that the 
poorly modelled opacities of FeH (Cushing
et al. 2003), as well as VO and TiO (McGovern et al. 2004)
also play an important role in this region.

\begin{figure}
\begin{center}
\includegraphics [width=65mm,angle=90]
{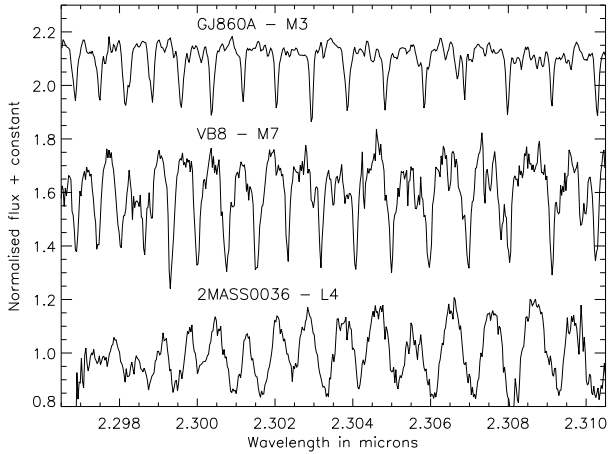}
\end{center}
\caption{Spectral sequence of CO bands from M3 to L4.}
\label{obsseq}
\end{figure}

\begin{figure}
\begin{center}
\includegraphics [width=70mm,angle=90]
{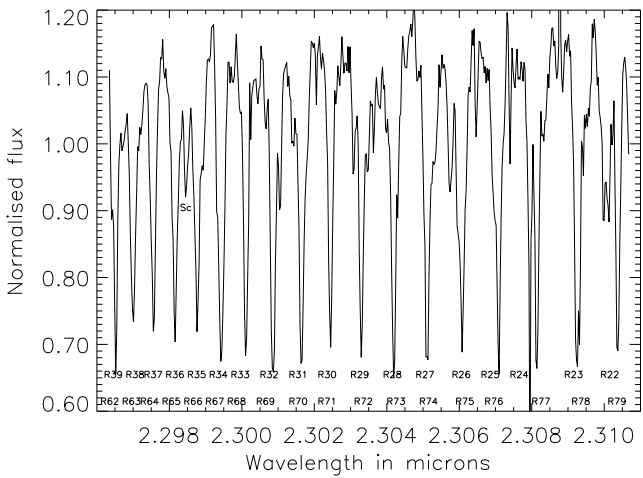}
\end{center}
\caption{CO 3-1 transitions and Sc line identified in GJ752B. The resolution of the data is insufficient to unambiguously identify the
higher energy (R62 to R79). The R24 and R77 transistions appear to be
resolved from one another, however, inspection of Fig. 3 indicates
that the left-hand feature is due to water vapour.}
\label{coid}
\end{figure}

\begin{figure}
\begin{center}
\includegraphics [width=85mm,angle=0]
{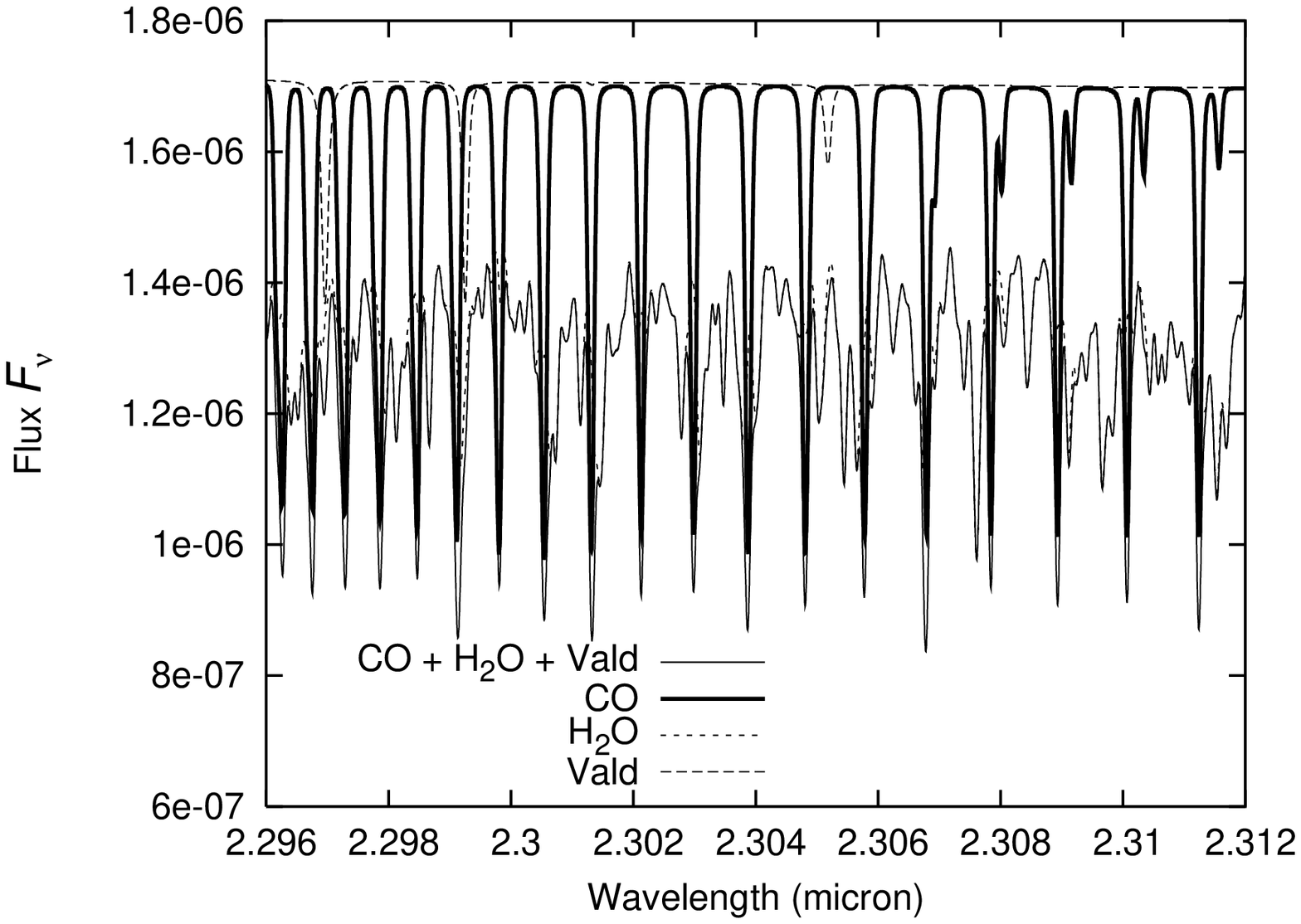}
\end{center}
\caption{The plot shows the various opacity contributions for a 3000~K log~g=5.0 
solar metallicity model from atomic lines (top of plot, dashed line), carbon 
monoxide transitions (top of the plot, thick solid line), water vapour (dotted line) 
and the overall formation of the continuum primarily dominated by water (thin solid  
line).}
\label{opacities}
\end{figure}

\begin{figure}
\begin{center}
\includegraphics [width=60mm,angle=0]{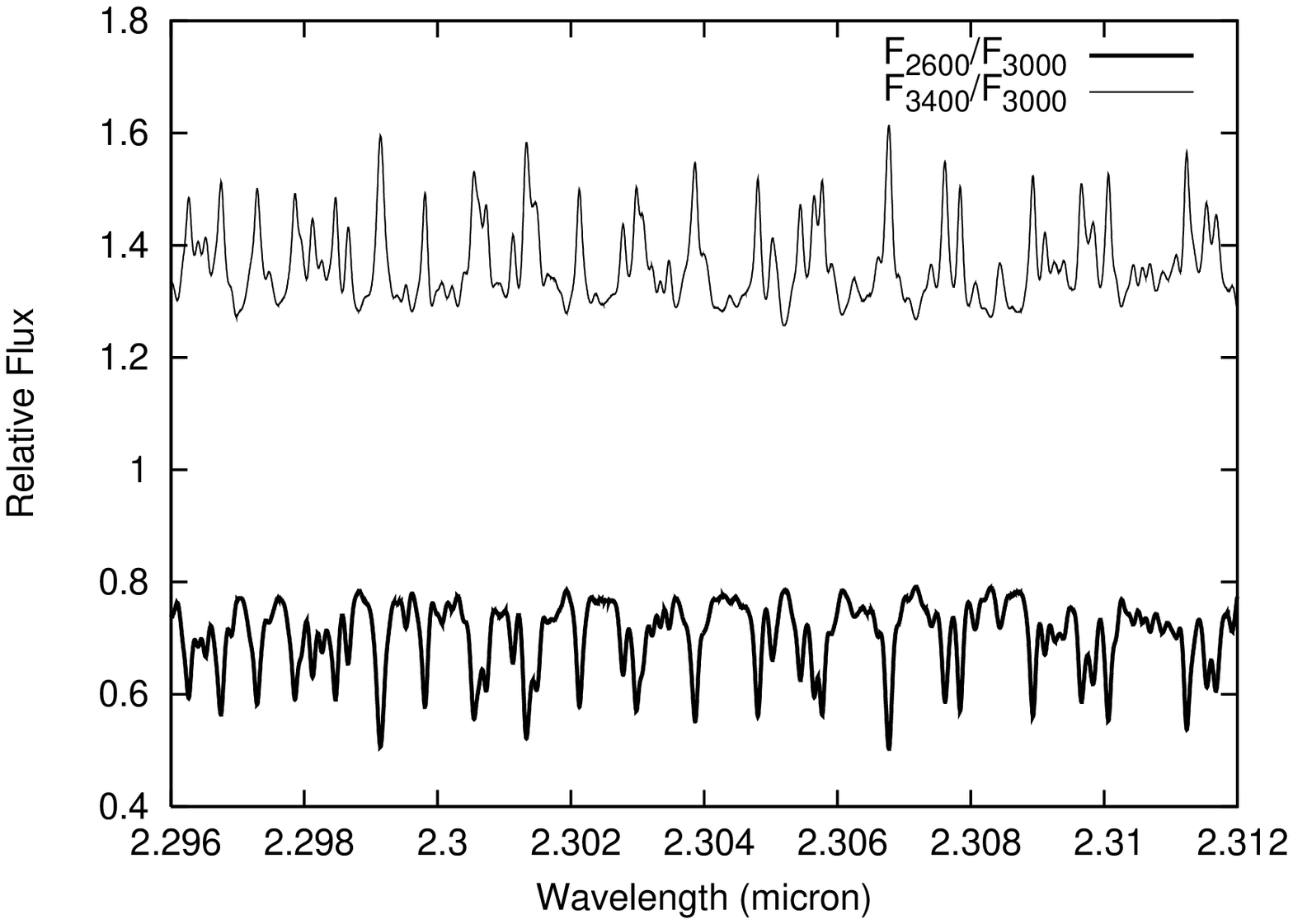}
\includegraphics [width=60mm,angle=0]{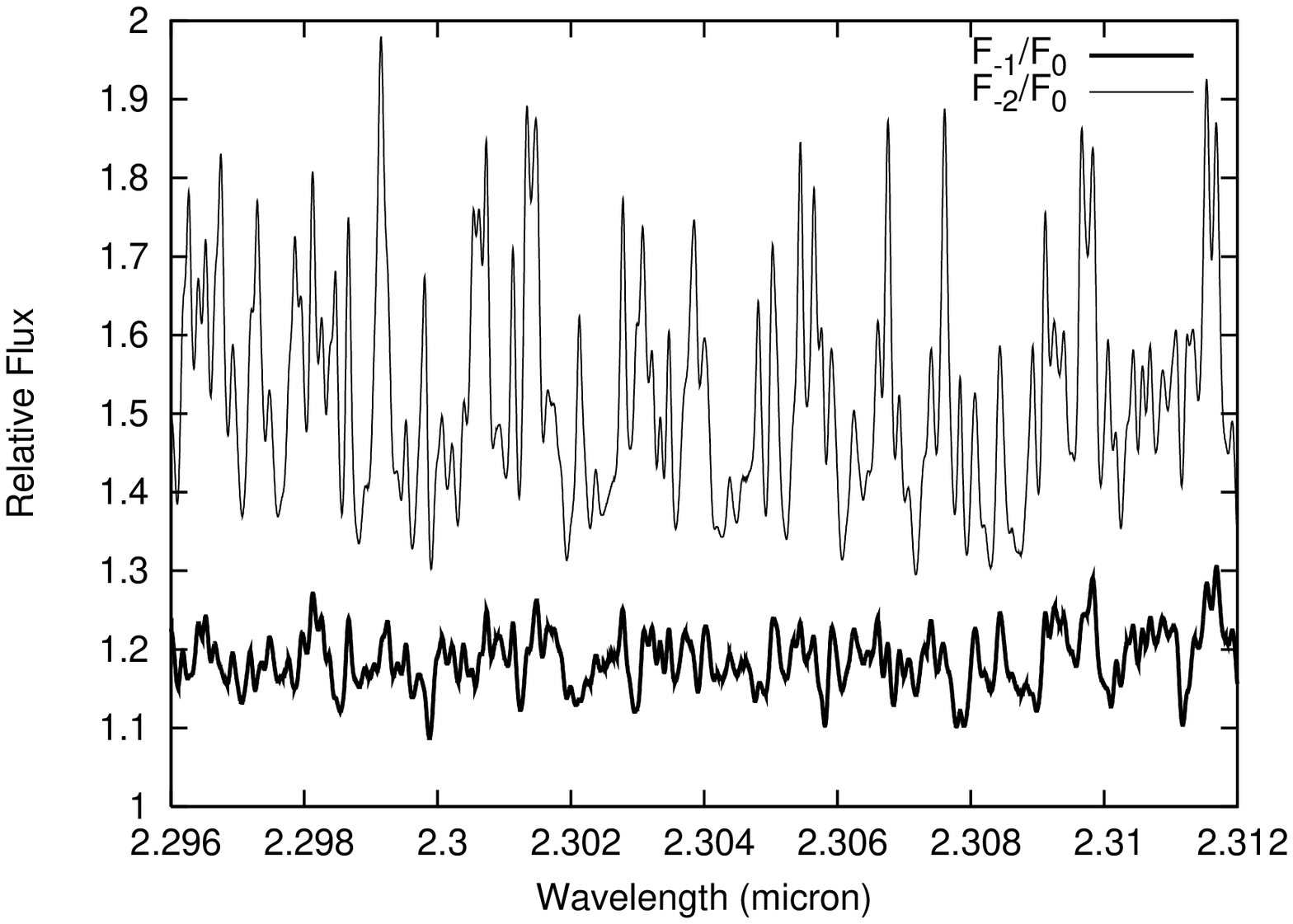}
\includegraphics [width=60mm,angle=0]{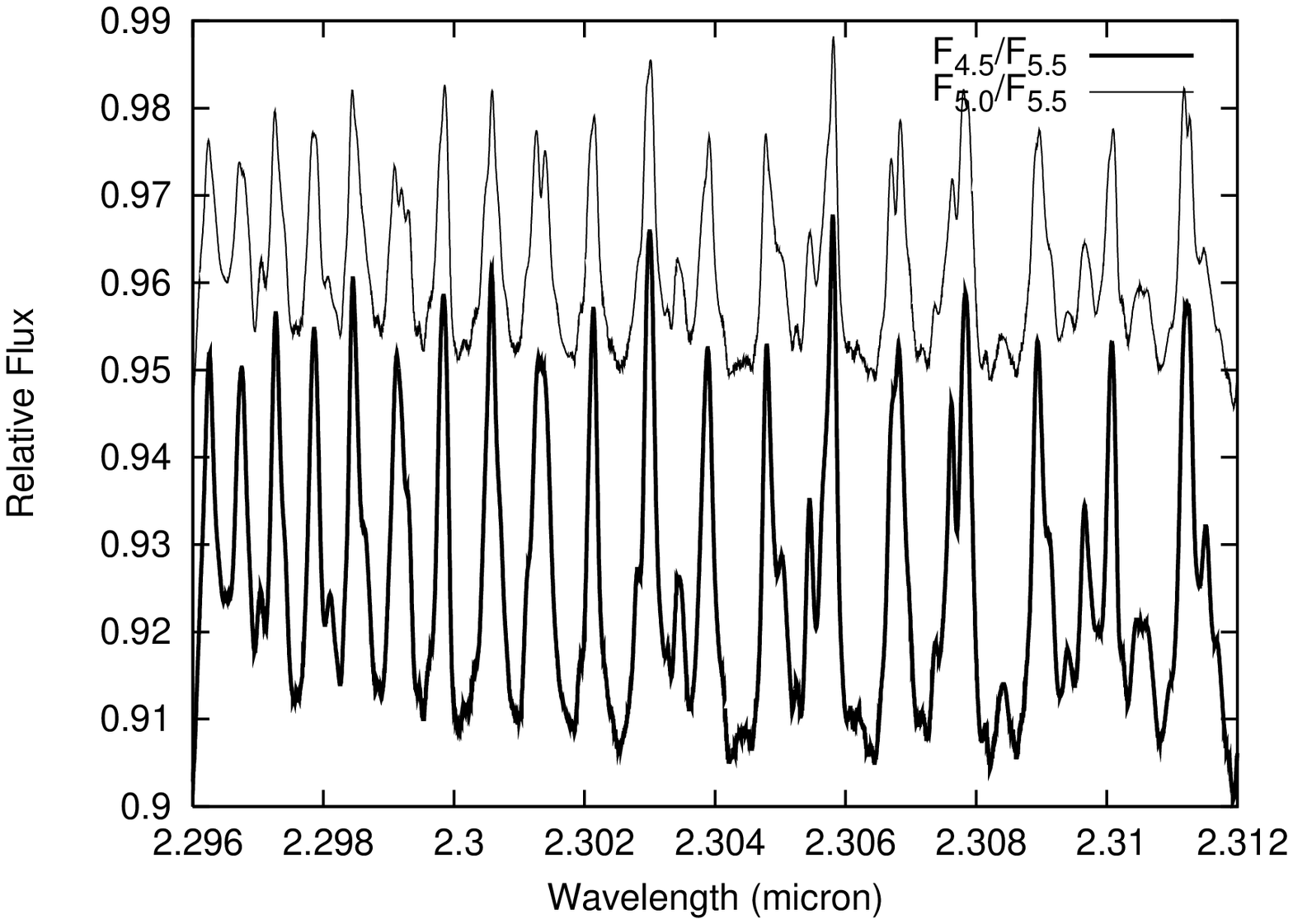}
\end{center}
\caption{The above plots show the temperature (top), metallicity (middle) and 
gravity (bottom) dependence of synthetic spectra around a base model of 3000~K, solar metallicity,
log g = 5.0. The value of the model atmosphere parameter being adjusted 
is indicated as subscript. For example in the middle plot, the label
F$_{-1}$/F$_{0}$ indicates the flux (F) of a 3000 K, [M/H]=--1, log g = 5.0 model
divided by a 3000 K, [M/H]=0, log g = 5.0 model.}
\label{sensitivity}
\end{figure}

\par
Here we investigate an alternative wavelength regime. 
In the spectral region between 2.29 and 2.45 $\mu$m, CO is a key
opacity for low mass stars.  CO appears in a relatively easily-observed
stable part of the K band and molecular data, including $f$-values,
are well known. Moreover, CO is believed to be formed under LTE
(e.g. Carbon et al. 1976) and therefore the levels are populated
according to the Boltzmann distribution. The available CO line
list has proven to be reliable for solar work and so is believed
to be more than adequate for the lower energy states accessed in cool 
dwarf atmospheres (Goorvitch 1994). The other significant metal 
diatomic species appearing in infrared cool dwarf stars are FeH, VO
and TiO, however, these diatomics are not as prominent as CO 
at wavelengths obtainable with the infrared echelle used for these
observations.

In Pavlenko \& Jones (2002) we showed that the $\Delta\nu$=2 carbon 
monoxide bands around 2.3--2.4~$\mu$m can be well modelled by synthetic 
spectra. This region is dominated by CO and H$_2$O bands, and has few atomic 
lines of significance. This is advantageous because CO is well modelled 
relative to the current quality of atomic oscillator strengths in the 
infrared (e.g. Jones et al. 1996, Lyubchik et al. 2004). Here we extend 
this work to much higher resolution, where the CO bands are very distinct 
from the water vapour modelled continuum. The wavelength range (2.297 to 2.311 
microns) was chosen on the basis of features in late type dwarf spectra 
identified to be relatively metallicity sensitive and reproducible by 
synthetic spectra in Viti et al. (2002).

\begin {table*}
\caption {Literature properties of observed targets: kinematic classification 
(KIN-class) are from Leggett (1992) and Leggett et al. (1998), spectral types 
are from Kirkpatrick, Henry \& McCarthy (1991) and Gizis (2002) and 
empirical temperatures derived using Lane et al. (2001), Segransan et al. (2003), 
Dahn et al. (2002) and Vrba et al. (2004).}
\begin {tabular} {ccccccccc}
Object & KIN-class & Sp Type & 
Empirical temperature (K) 
\\\\\\
GJ860A     &  OD & dM3   & 3310 \\
GJ725A     &  OD & dM3   & 3310 \\
GJ725B     &  OD & dM3.5 & 3230 \\
GJ896A     &  YD & dM3.5 & 3230 \\
G87-9B     &  -- & dM4   & 3150 \\
GJ699      & O/H & dM4   & 3150 \\
GJ860B     &  OD & dM4   & 3150 \\
GJ896B     &  YD & dM4.5 & 3070 \\
GJ630.1A   &   H & dM4.5 & 3070 \\
GJ166C     &  OD & dM4.5 & 3070 \\
GJ2005ABCD &  OD & dM5.5 & 2910 \\
GJ65A      &  YD & dM5.5 & 2910 \\
GJ65B      &  YD & dM6   & 2825 \\
GJ644C     &  OD & dM7   & 2670 \\
GJ752B     &  OD & dM8   & 2550 \\
LP944-20   &  YD & dM9   & 2440 \\
2MASS0036  & - &   dL4   & 1900 \\
\end{tabular}
\end{table*}

\begin {table*}
\caption {Derived synthetic spectra parameters are given. The syntax for the 
models is \Tef/~log g~/~[M/H] so for example 2800/5.0/-0.5 means a 2800~K,
gravity = 5.0 cm/s and metallicity --0.5 dex model. Rotational velocities
are derived from the unconstrained minimisation fit and are 
typically accurate to 3 km/s. Values given in italic are fixed. The values 
for the minimisation of coolest objects LP944-20 and 2MASS0036 (shortened
its full designation of 2MASS J00361617+1821104) are rather 
dependent on the details of the model. Minimisations for these objects are 
presented in Fig. \ref{ldwarfs}.}
\begin {tabular} {ccccccccc}

Object & Rotational velocity& unconstrained & [M/H] constrained & 
\Tef constrained \\
& v~sin~i (km/s) & minimisation & \Tef \& log~$g$ minimisation & 
[M/H] \& log~$g$ minimisation\\\\\\
GJ860A     &  7.0 & 3500/5.5/0.0 & 3500/5.5/$\it{0.0}$ & $\it{3300}$/5.0/-1.0 \\
GJ725A     &  5.0 & 3500/5.5/-0.5 & 3600/5.5/$\it{0.0}$ & $\it{3300}$/5.0/-1.5 \\
GJ725B     &  7.0 & 3500/5.5/0.0 & 3500/5.5/$\it{0.0}$ & $\it{3200}$/5.0/-1.5 \\
GJ896A     & 10.0 & 3400/4.5/-1.5 & 3600/5.5/$\it{0.0}$ & $\it{3200}$/4.5/-2.0 \\
G87-9B     &  6.0 & 3500/4.5/-0.5 & 3500/5.5/$\it{0.0}$ & $\it{3200}$/5.0/-1.5 \\
GJ699      &  5.0 & 3500/4.5/-0.5 & 3600/5.5/$\it{0.0}$ & $\it{3200}$/4.5/-2.0 \\
GJ860B     &  8.0 & 3500/5.5/0.0 & 3500/5.5/$\it{0.0}$ & $\it{3200}$/4.5/-1.5 \\
GJ896B     & 15.0 & 3500/5.0/-0.5 & 3500/5.5/$\it{0.0}$ & $\it{3100}$/4.5/-2.0 \\
GJ630.1A   & 27.5 & 3300/5.0/-0.5 & 3300/5.0/$\it{0.0}$ & $\it{3100}$/4.5/0.0 \\
GJ166C     &  5.0 & 3400/5.5/-0.5 & 3500/5.5/$\it{0.0}$ & $\it{3000}$/4.5/-2.0 \\
GJ2005ABCD & 14.0 & 3000/5.5/-1.0 & 3300/4.5/$\it{0.0}$ & $\it{2900}$/5.0/-1.5 \\
GJ65A      & 31.5 & 3400/4.5/0.0 & 3400/4.5/$\it{0.0}$ & $\it{2900}$/5.5/-1.5 \\
GJ65B      & 29.5 & 3300/4.5/0.0 & 3300/4.5/$\it{0.0}$ & $\it{2800}$/5.5/-1.5 \\
GJ644C     & 12.5 & 2900/4.5/0.0 & 2900/4.5/$\it{0.0}$ & $\it{2700}$/5.0/-1.0 \\
GJ752B     & 10.5 & 2900/5.5/0.0 & 2900/5.5/$\it{0.0}$ & $\it{2600}$/5.0/-1.5 \\
LP944-30   & 31.0 & -- & -- & -- \\
2MASS0036  & 38.0 & -- & -- & -- \\
\end{tabular}
\end{table*}

\section{Observations}
The targets chosen for this study are all bright relatively well-studied M and 
L dwarfs. The source selection was made in order to give good coverage in spectral 
type and metallicity. However, the half nights available to us limited our sample 
to a relatively restricted range of right ascension. The sample is shown in Table 1.

The targets were observed during the first half of the nights of 2001 
September 8--12 with the Cooled Grating Spectrometer 4 (CGS4) on the UK 
Infrared Telescope (UKIRT) on Mauna Kea, Hawaii. The weather was photometric 
throughout with optical seeing of typically 0.8". Comparison sky spectra 
were obtained by nodding the telescope so that the object was measured 
successively in two rows of the array, separated by 30 arcsec.

\par The echelle grating in 24th order at a central wavelength setting of 
2.304 microns was used for all observations. This setup gives wavelength 
coverage from 2.297 to 2.311 microns at a resolution of approximately 42000.
\par To remove telluric bands of water, oxygen, carbon dioxide and methane, 
we observed A and B standard stars. Such stars are not expected to have 
features in common with cool dwarf stars and appear to be featureless across 
our spectral range. Wavelength calibration was carried out using a xenon arc 
lamp. This generally worked well because although there are only four lines 
available in this region, the xenon lines at 2.29 and 2.31 microns fall at 
either edge of the array and so provide a good wavelength calibration. The 
only caveat to this is that the wavelength positioning of the echelle 
is only accurate to around 
20 pixels and so the desired wavelength interval is sometimes shifted redward 
or blueward by around 0.001 microns. Our cross-correlation tests indicate 
that the wavelength calibration was better than 0.0001 microns. Sky subtraction 
was done with standard routines which take into account any residual sky 
emission due to variation of the sky brightness between paired object and sky 
spectra. The signal was spread between three rows. To extract the spectrum 
from the sky subtracted signal an Optimal Extraction technique was used; this 
combines the rows using weights based on the spatial profile of the stellar 
image.  The spectra were reduced using the $\sc{Figaro}$, $\sc{Specdre}$ and 
$\sc{Kappa}$ packages provided and supported by Starlink.

\par A spectral sequence from M3 to L4 is shown in Fig. \ref{obsseq}. It can be 
seen from Figure 2 that individual rotational CO transitions can be resolved in 
our observed spectra. The CO opacities in our spectra are made up of a large number 
of spectral lines, covering a wide intensity range, of the second overtone ($\nu$ 
= 2--0).  The second overtone band of CO originates from vibration-rotation 
transitions in the ground electronic state $X^1\Sigma$ and obeys the selection 
rules $\Delta\nu$ = 2 and $\Delta J$ = $\pm$1.  The band head of the second overtone 
(i.e the point at which the separation between the R transitions is zero), occurs 
at $\sim$ 2.290 $\mu$m and therefore both `hot' (such as R77) and `cold' (such as 
R24) rotational transitions are seen in our spectra.

\begin{figure}
\begin{center}
\includegraphics [width=60mm,angle=0]
{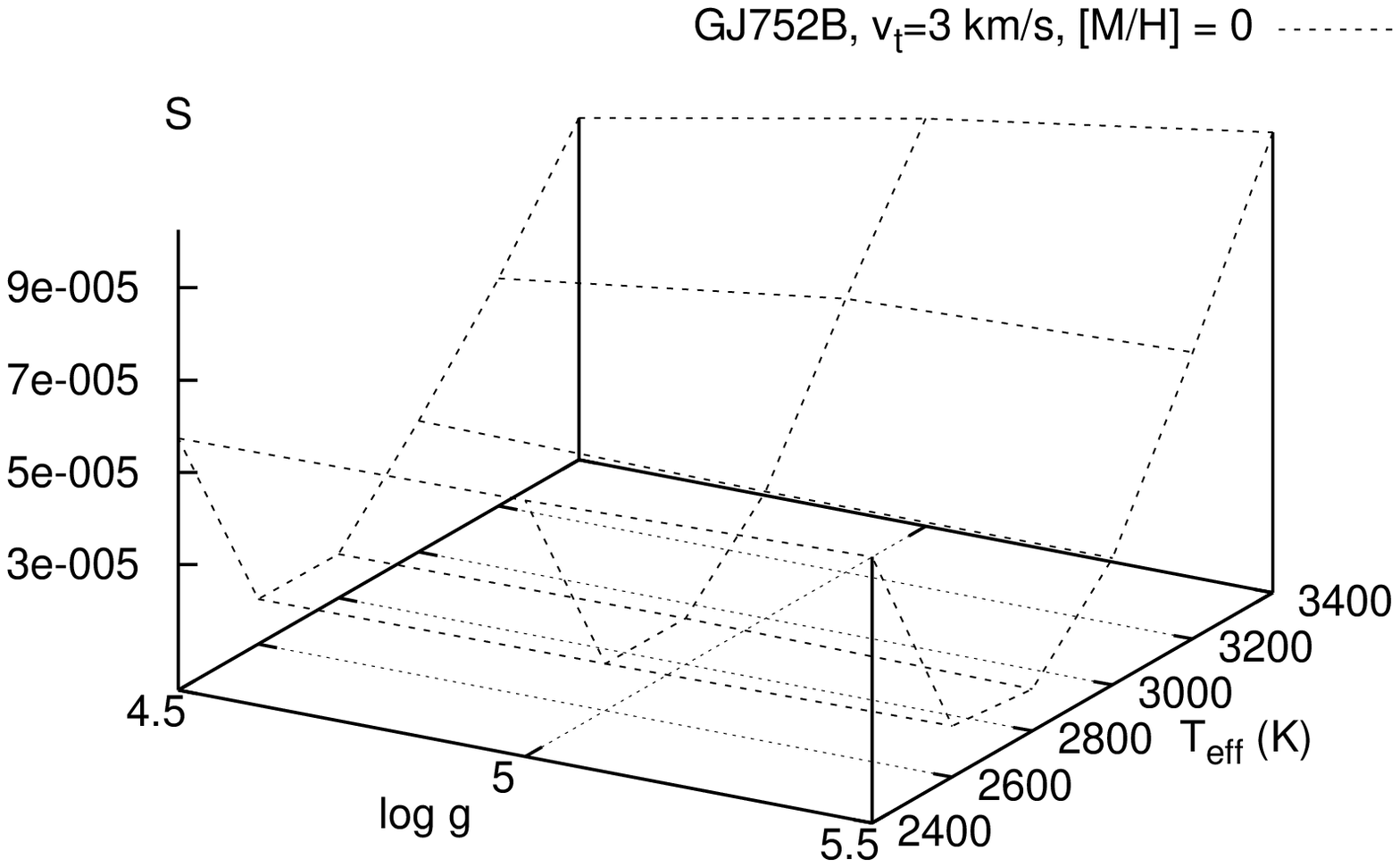}
\includegraphics [width=60mm,angle=0]
{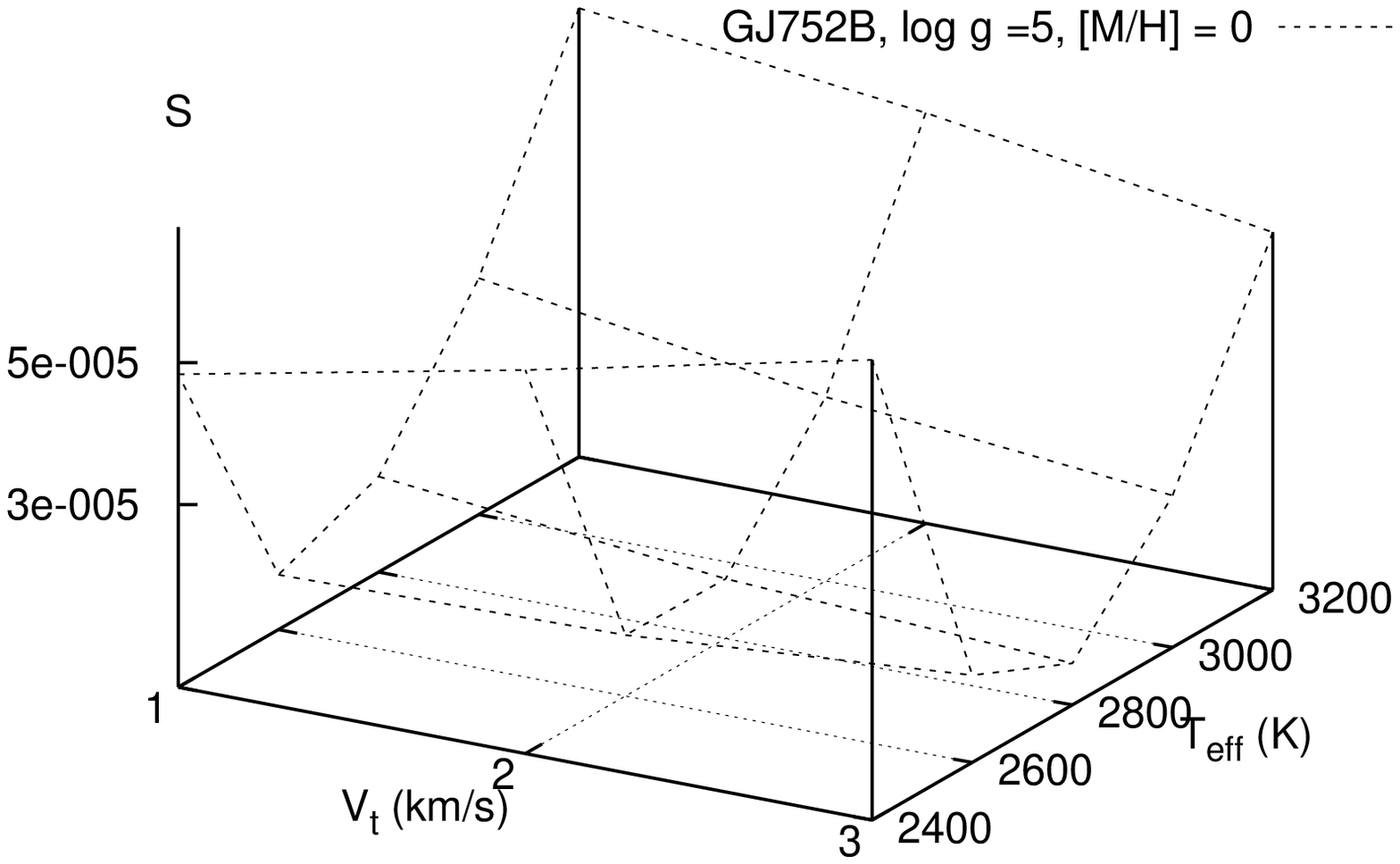}
\includegraphics [width=60mm,angle=0]
{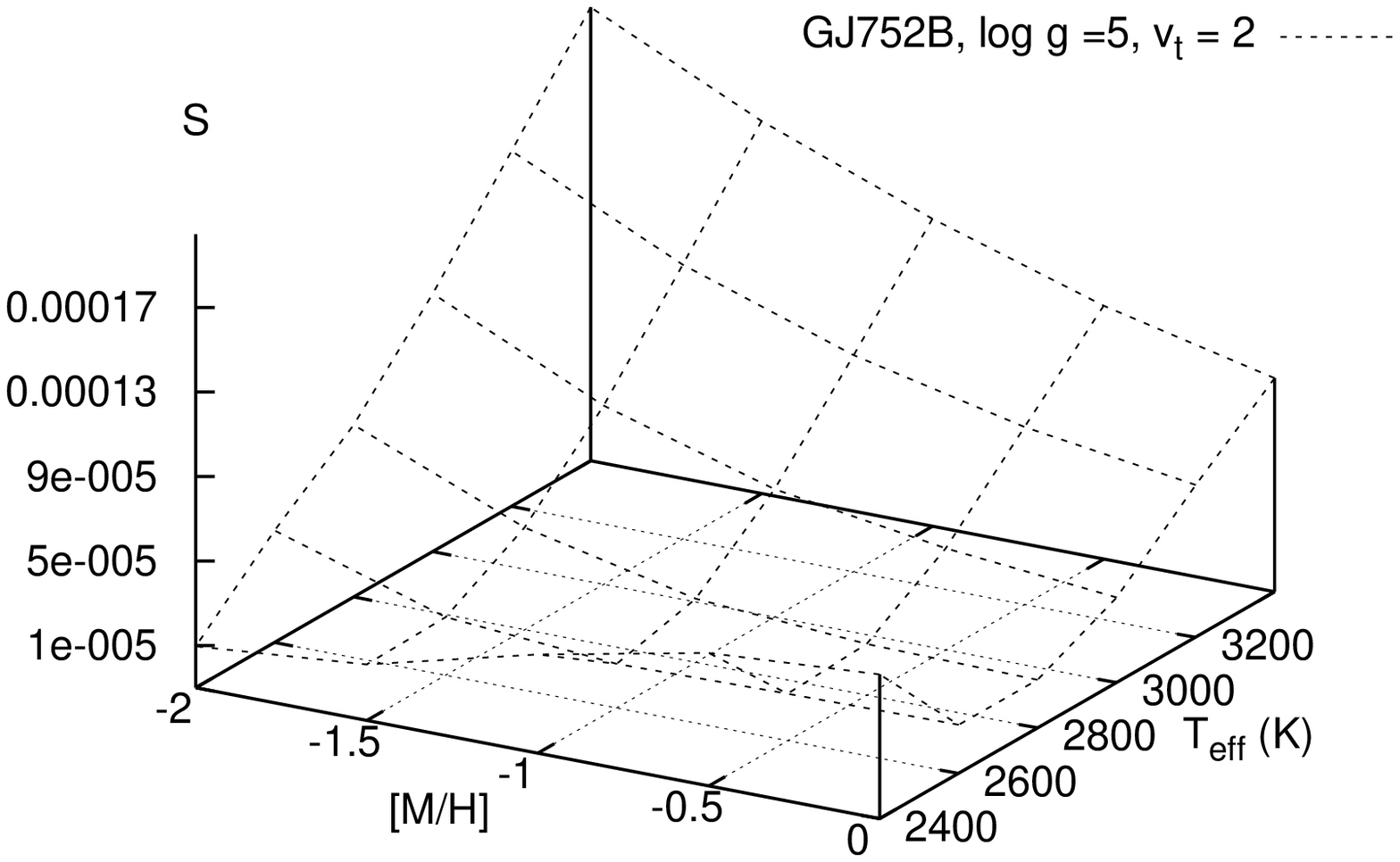}\end{center}
\caption{Surface plots showing the sensitivity of the best fit temperature
of GJ752B to metallicity, gravity and turbulent velocity. A base model of
solar metallicity, log $g$=5.0 and v$_t$=3km/s is used. }
\label{surface}
\end{figure}

\begin{figure*}
\begin{center}
\includegraphics [width=53mm,angle=0]{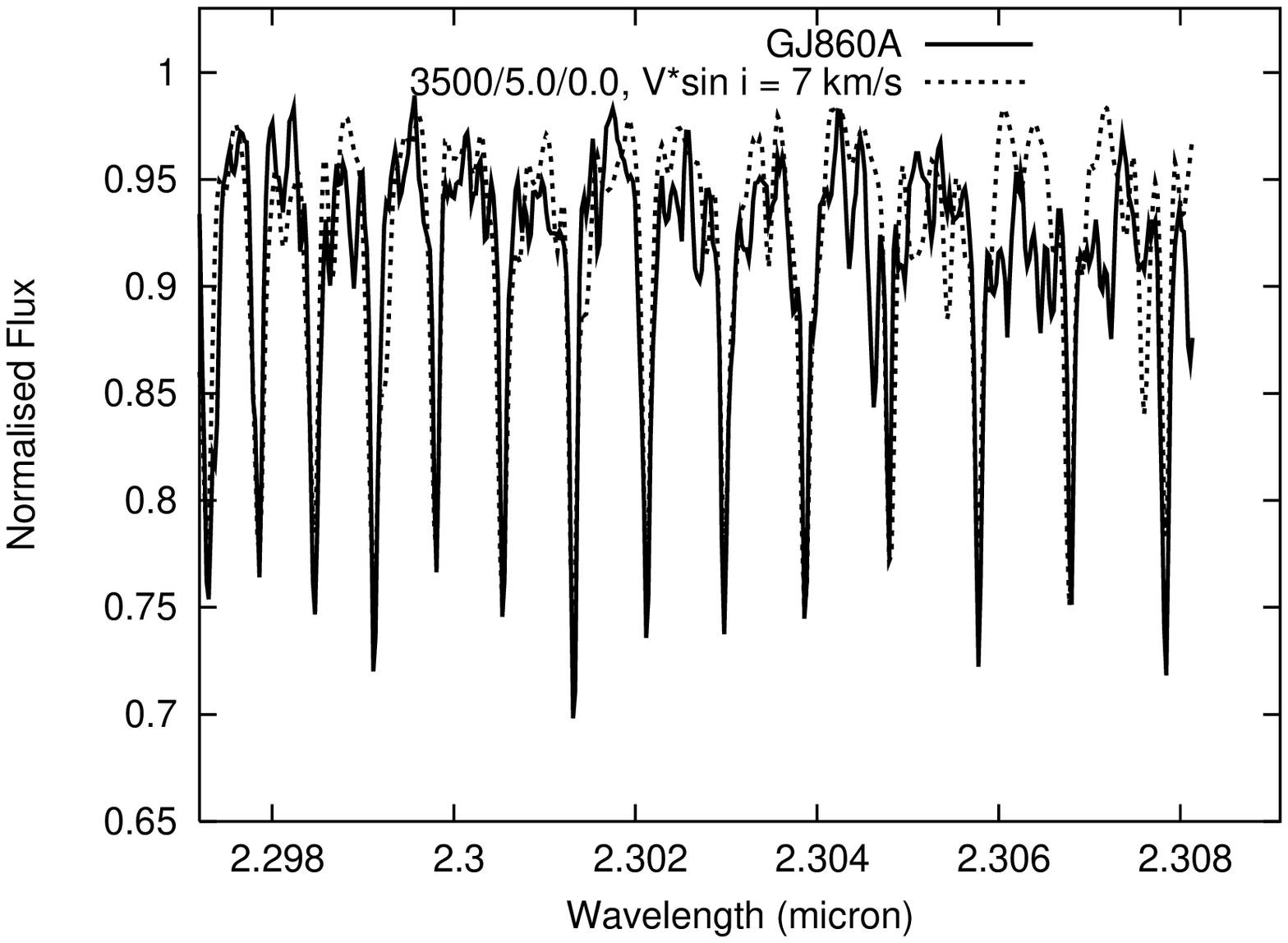}
\includegraphics [width=53mm,angle=0]{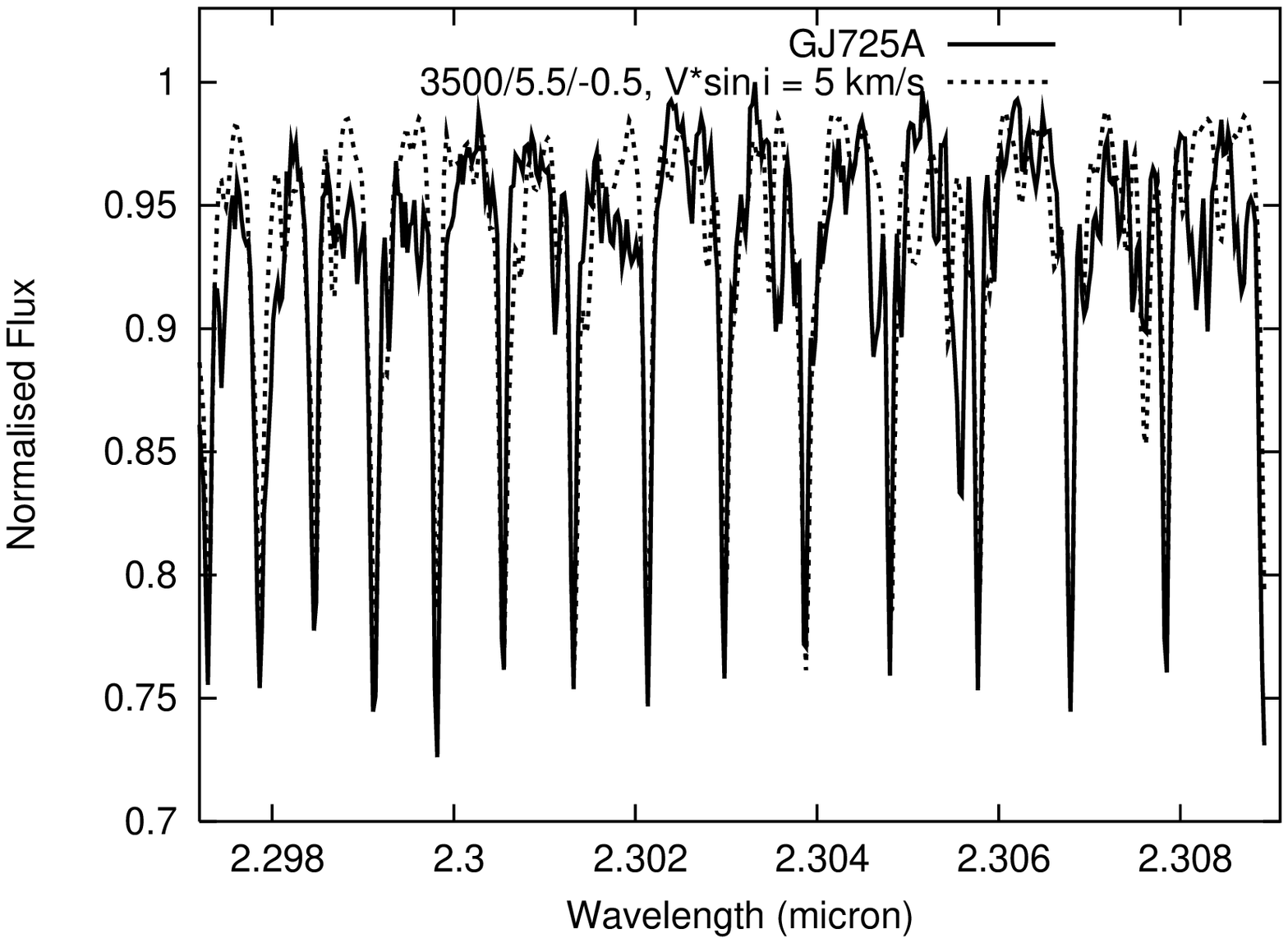}
\includegraphics [width=53mm,angle=0]{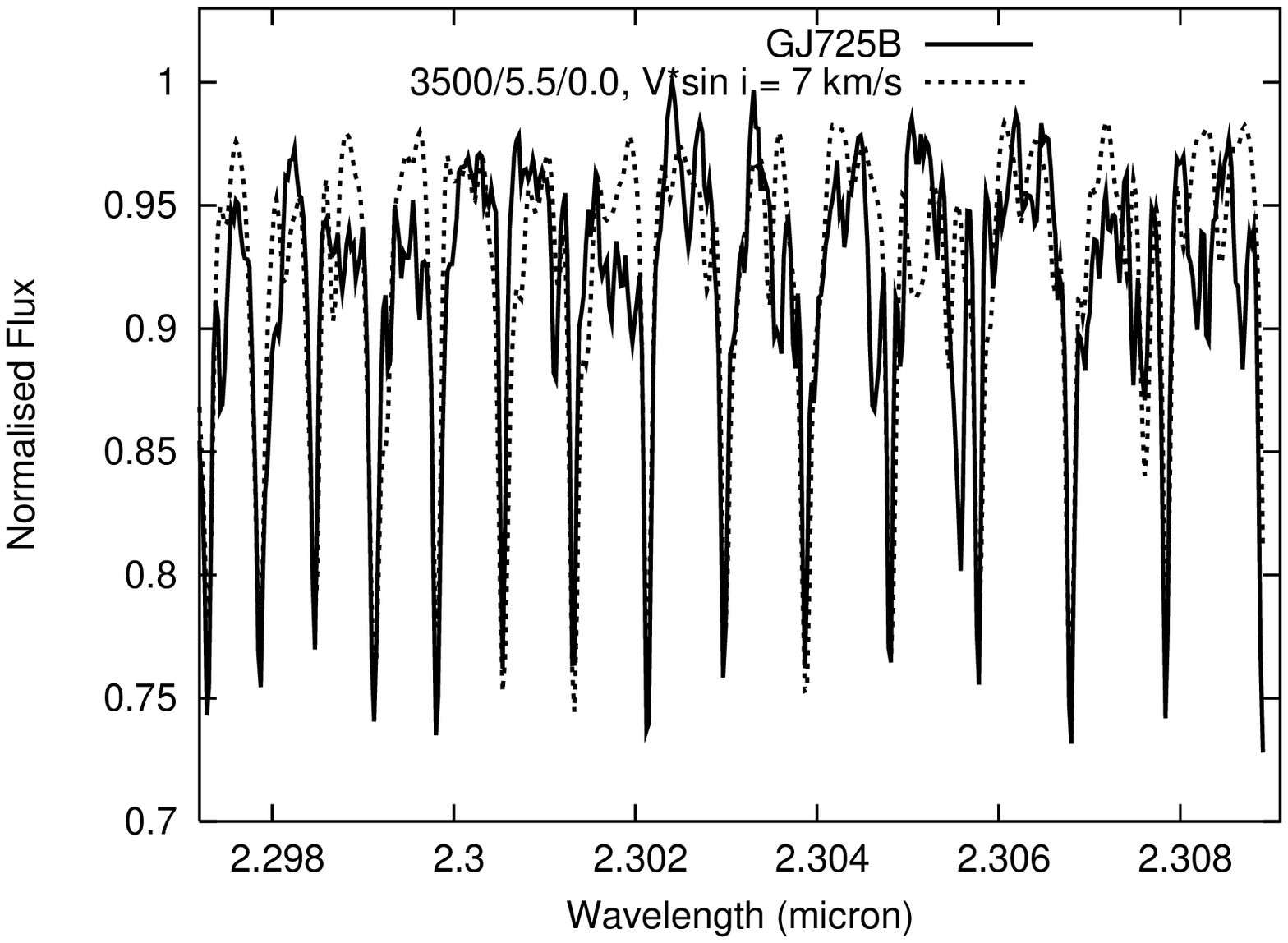}
\includegraphics [width=53mm,angle=0]{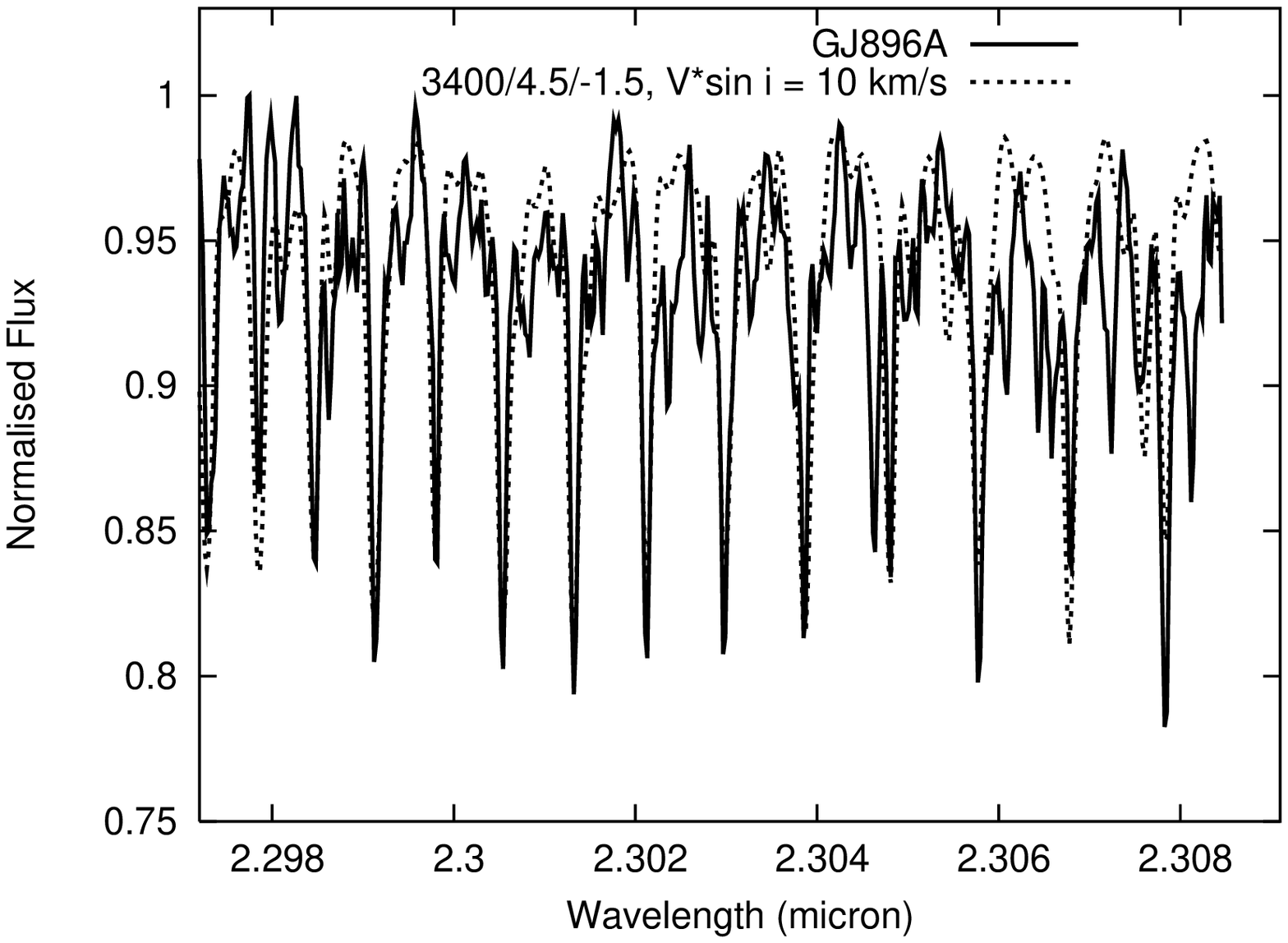}
\includegraphics [width=53mm,angle=0]{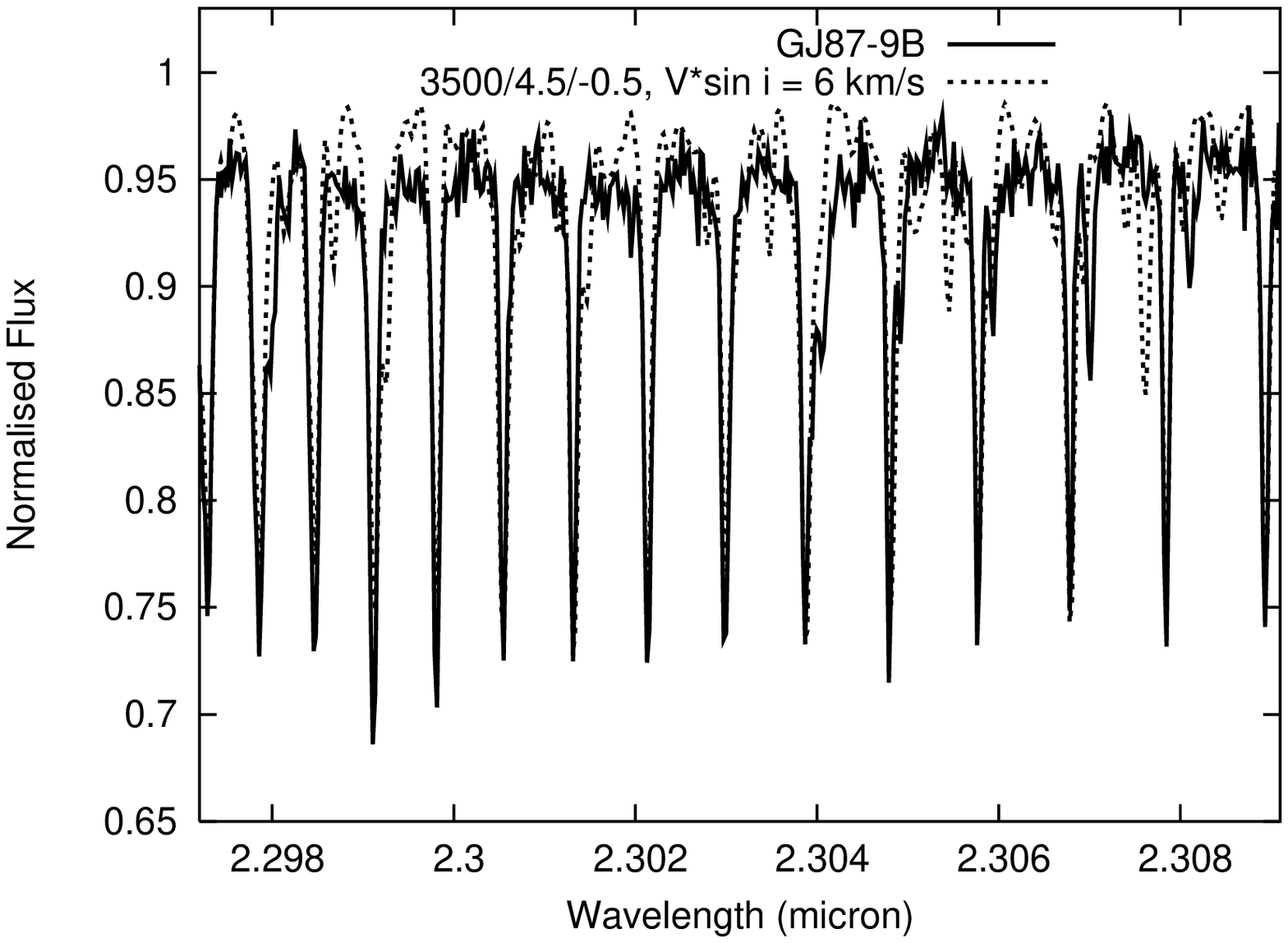}
\includegraphics [width=53mm,angle=0]{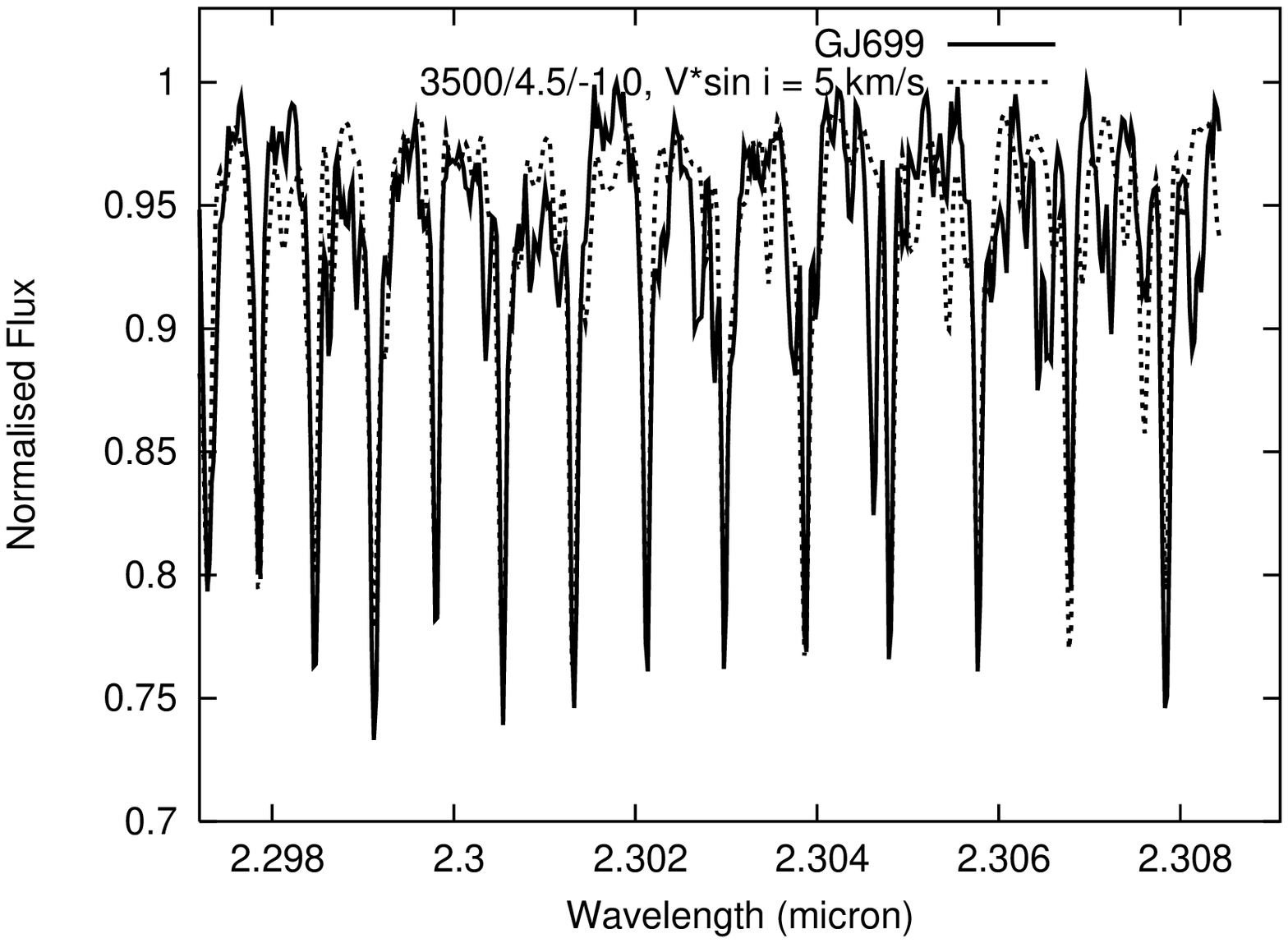}
\includegraphics [width=53mm,angle=0]{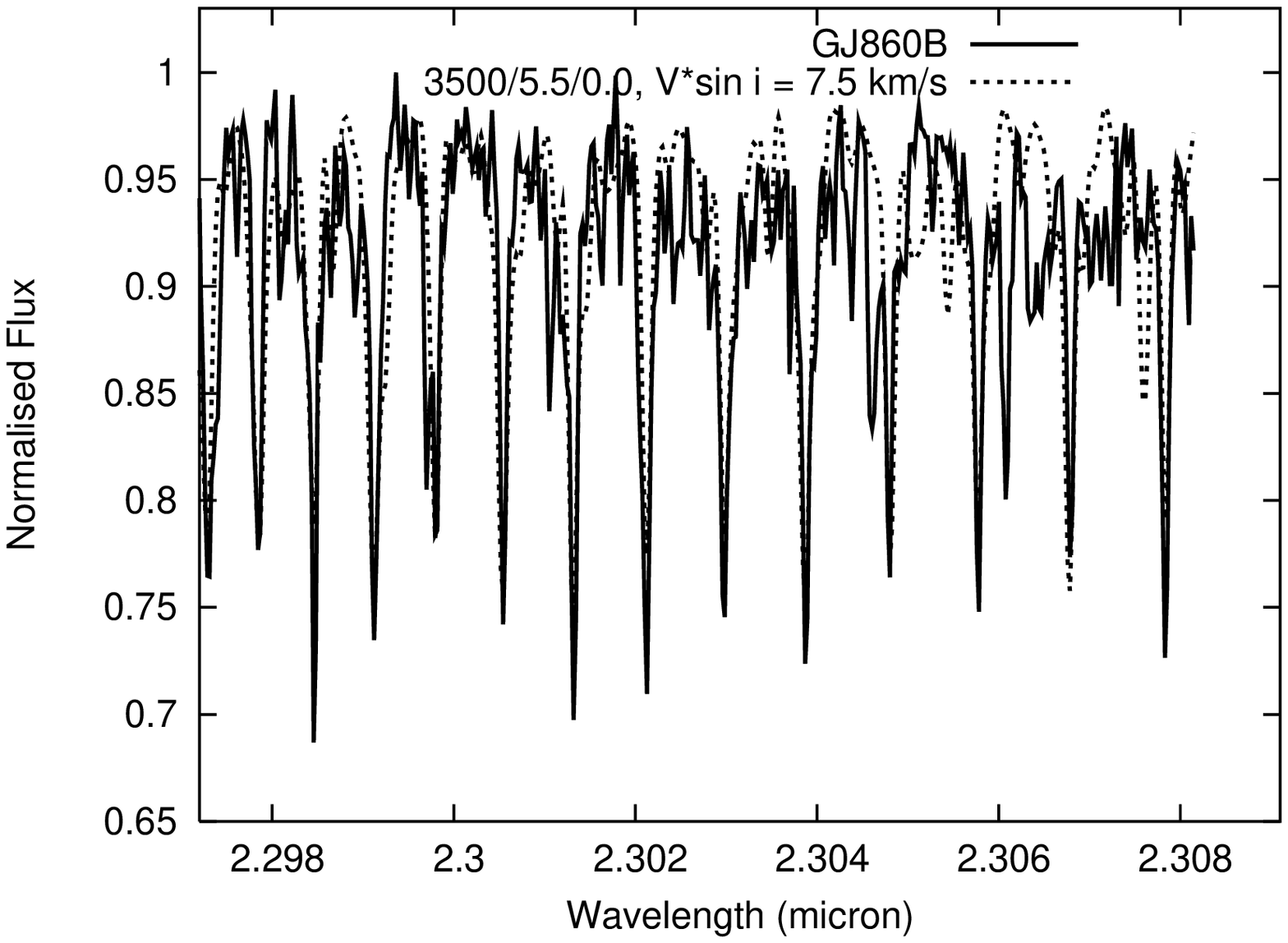}
\includegraphics [width=53mm,angle=0]{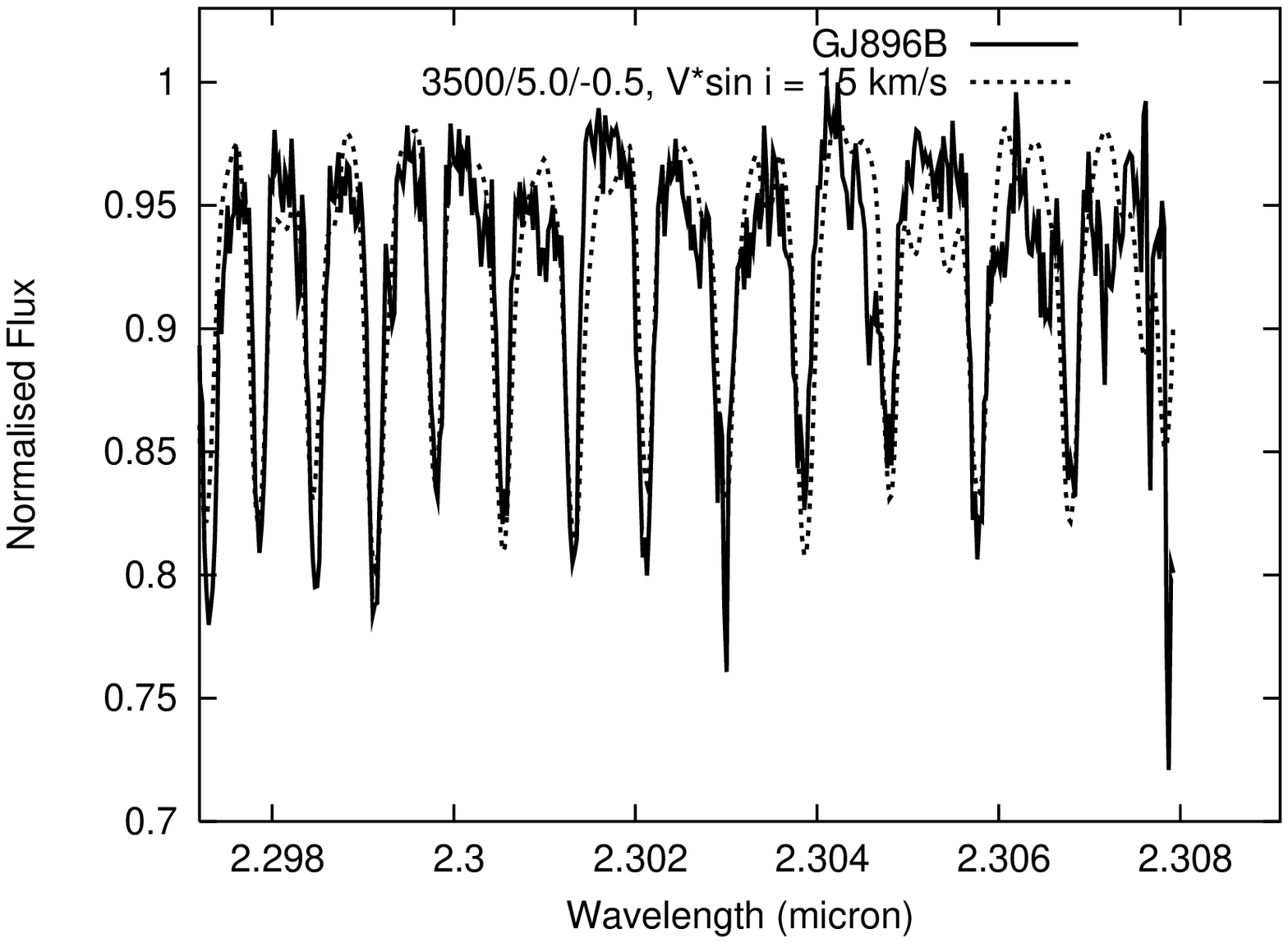}
\includegraphics [width=53mm,angle=0]{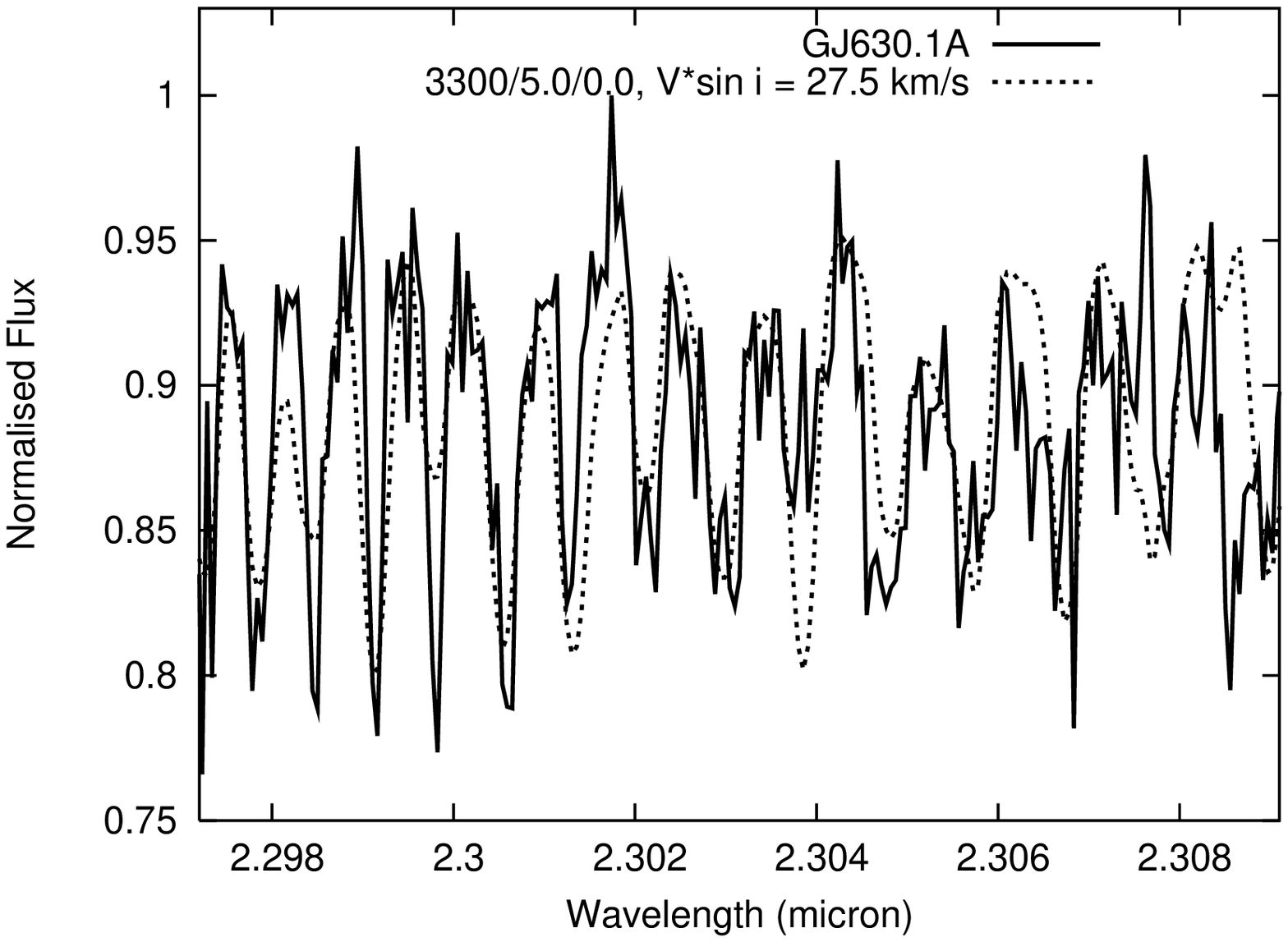}
\includegraphics [width=53mm,angle=0]{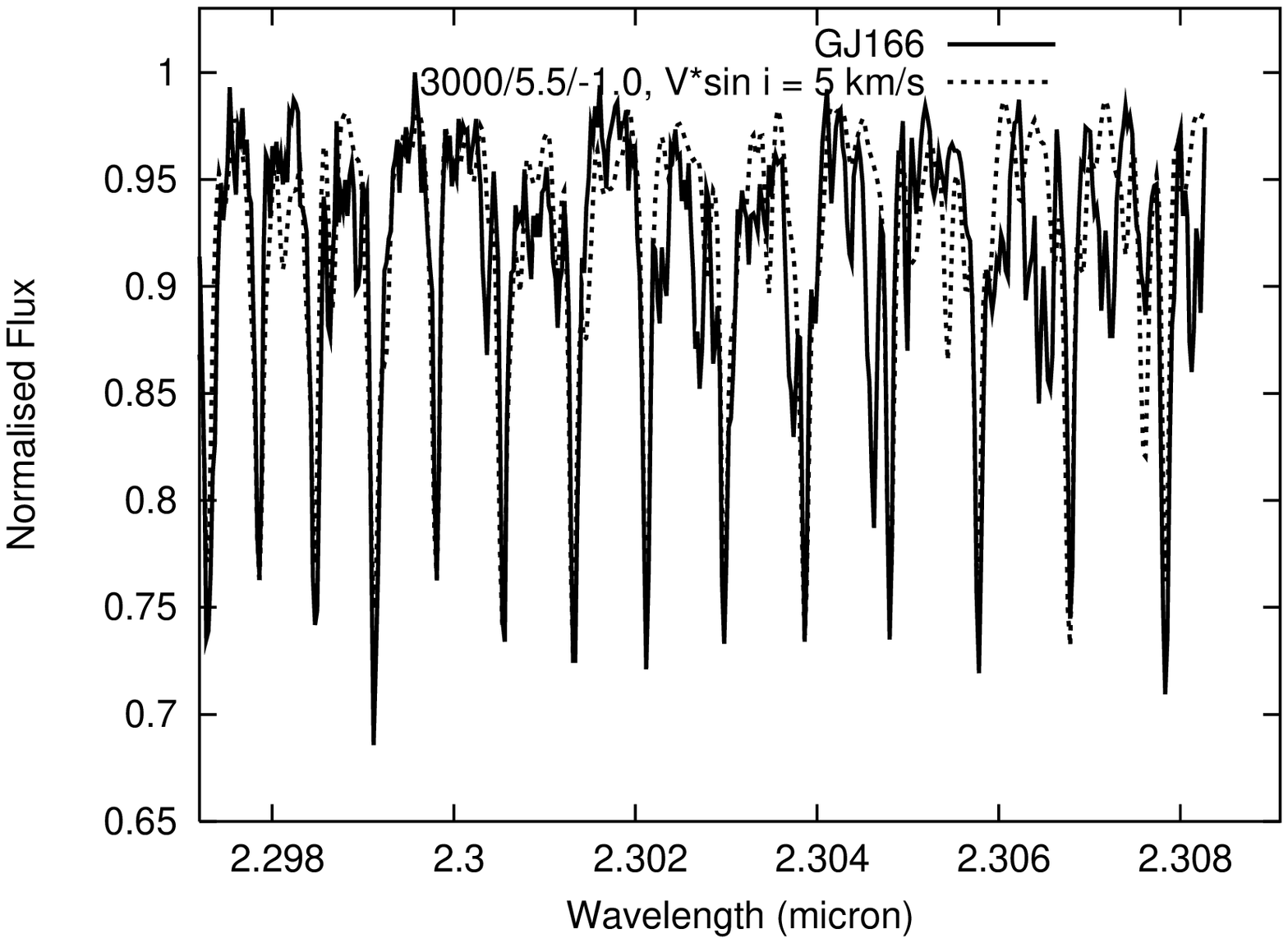}
\includegraphics [width=53mm,angle=0]{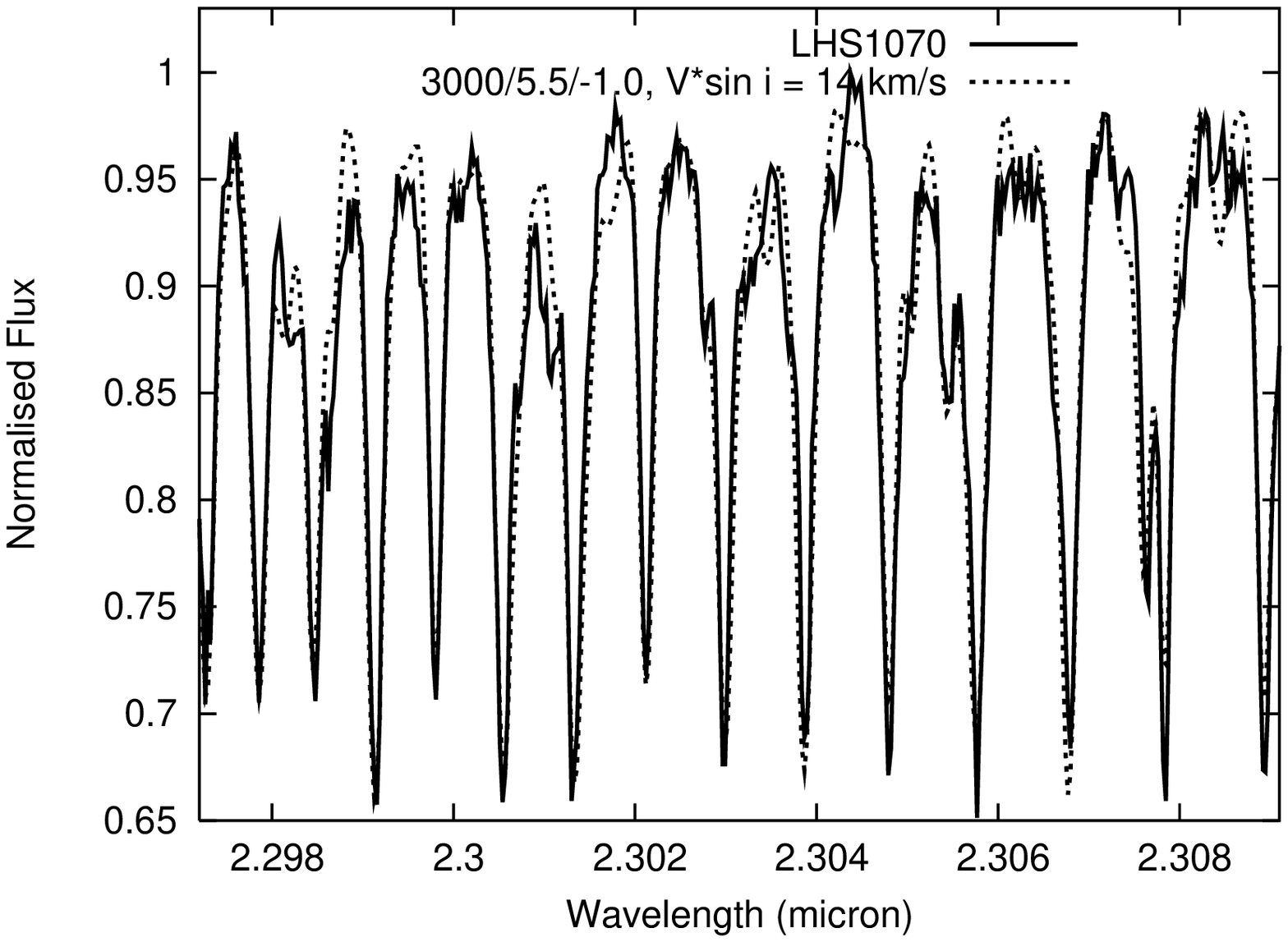}
\includegraphics [width=53mm,angle=0]{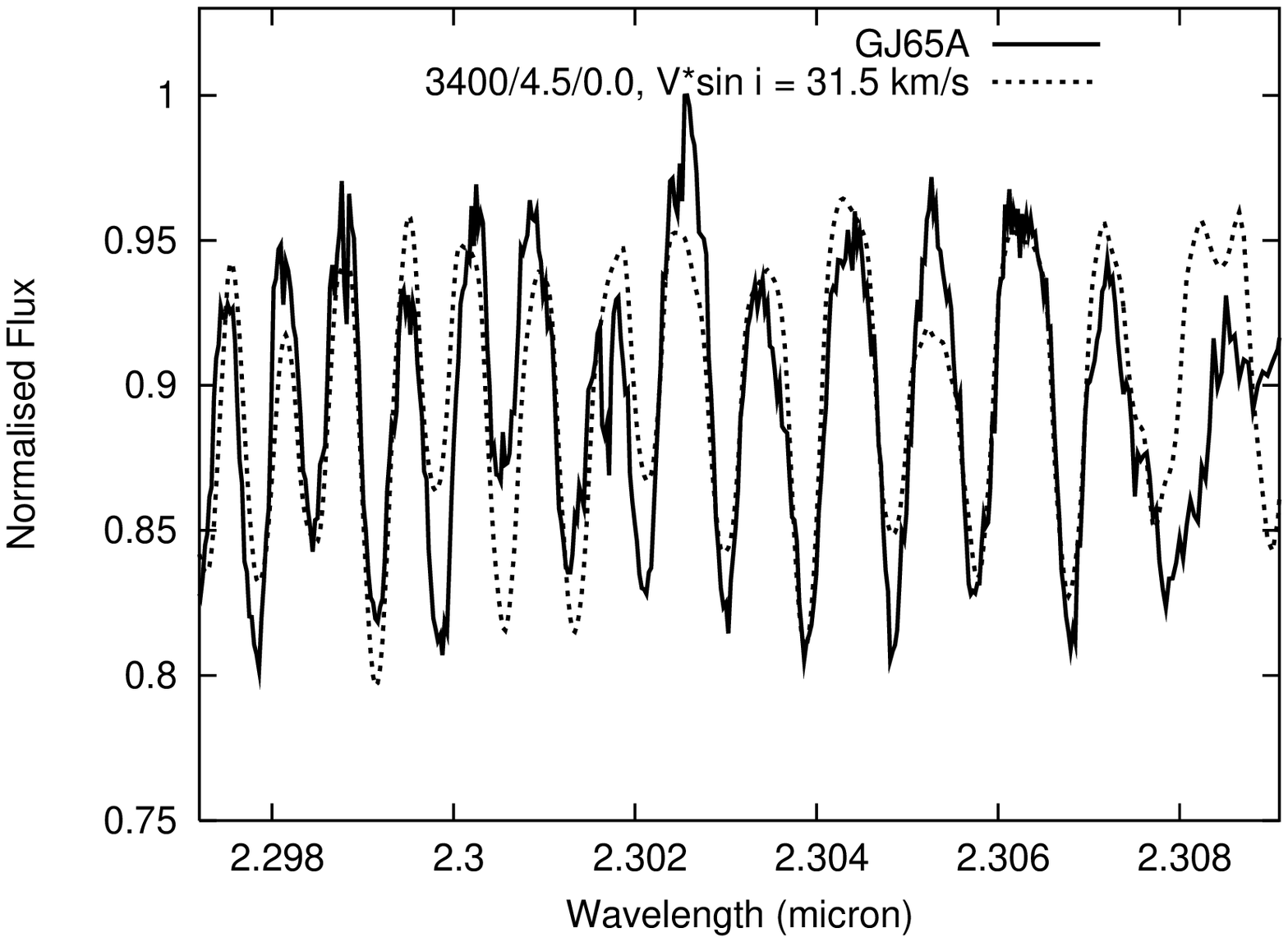}
\includegraphics [width=53mm,angle=0]{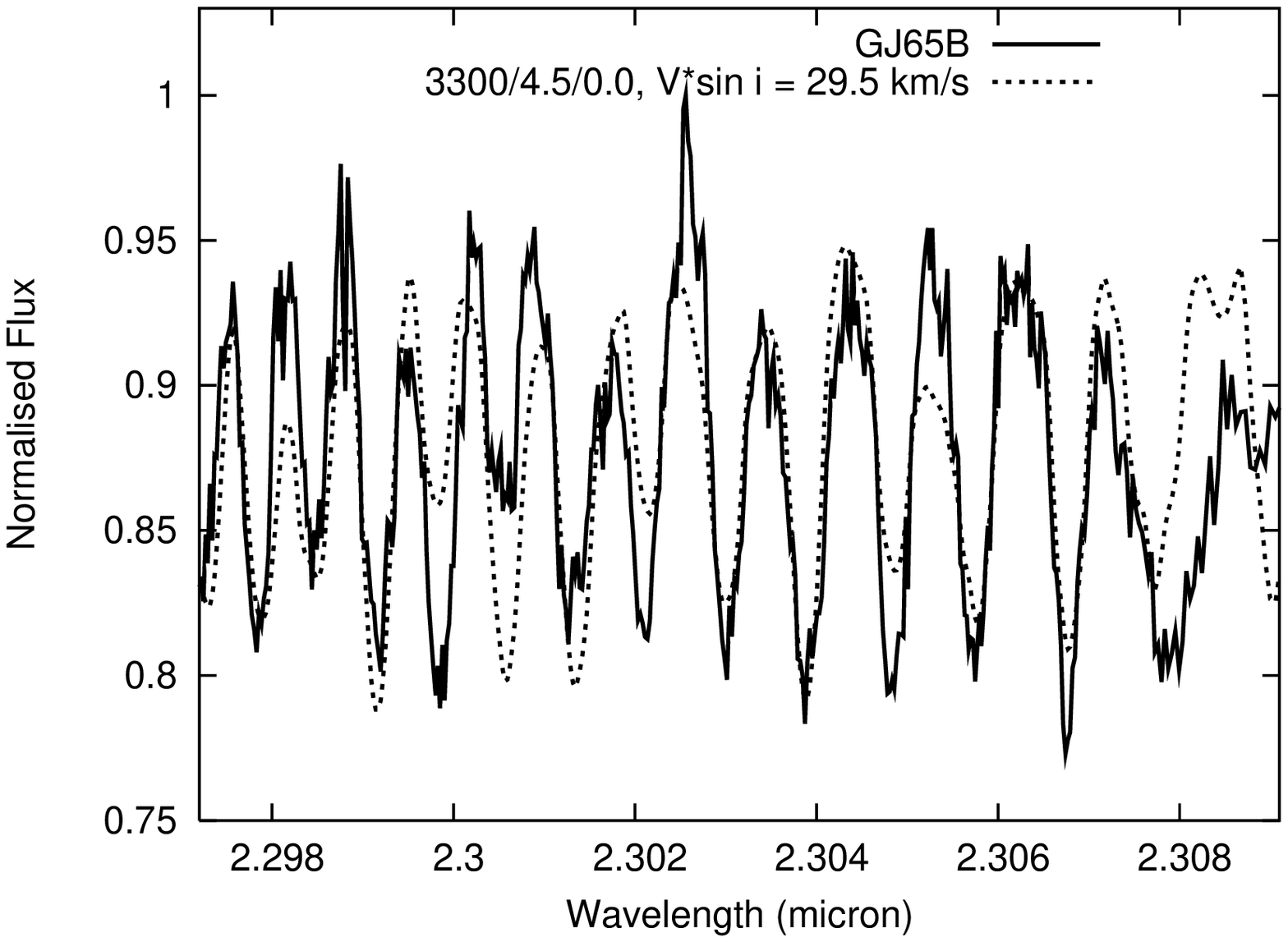}
\includegraphics [width=53mm,angle=0]{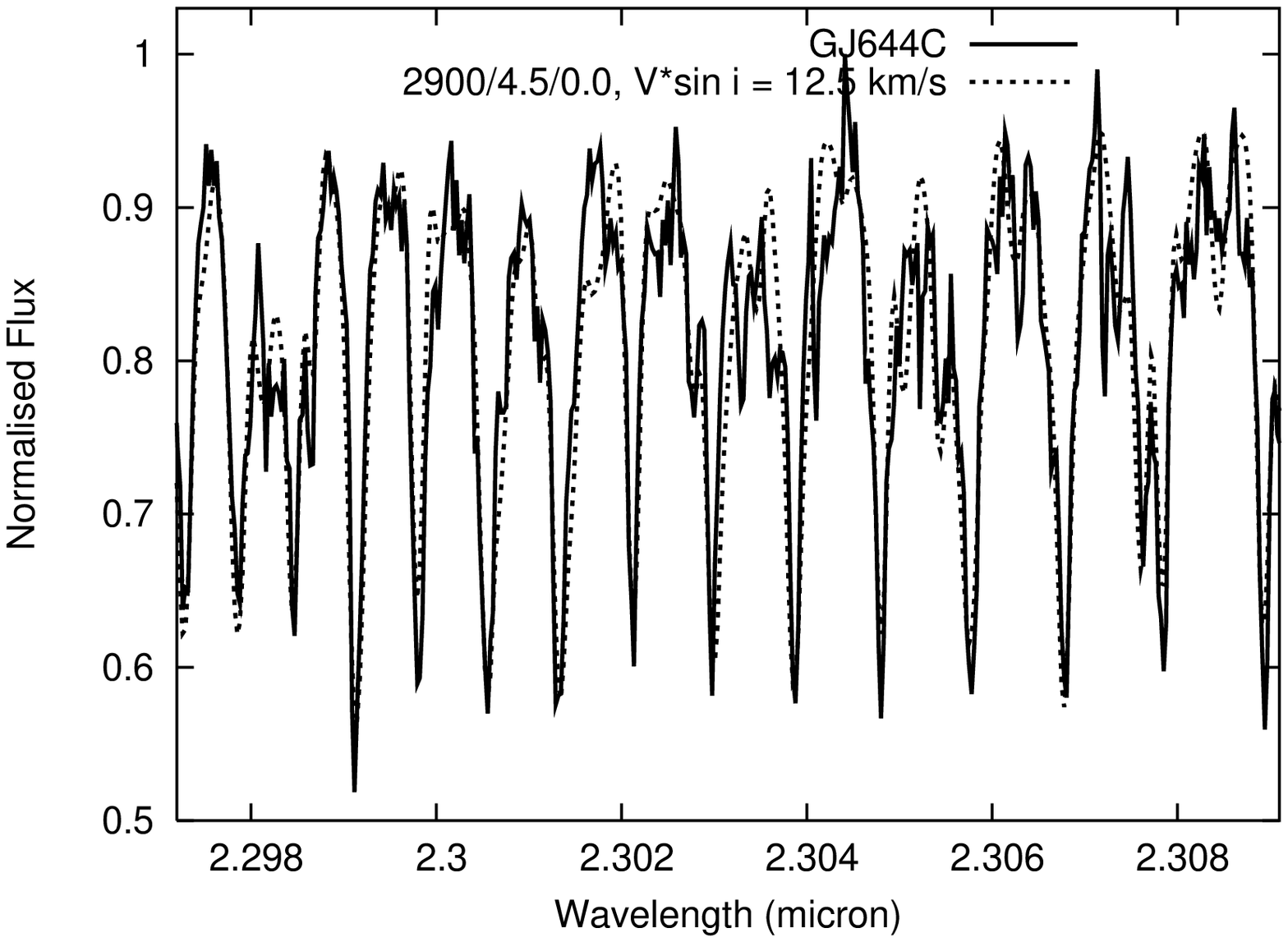}
\includegraphics [width=53mm,angle=0]{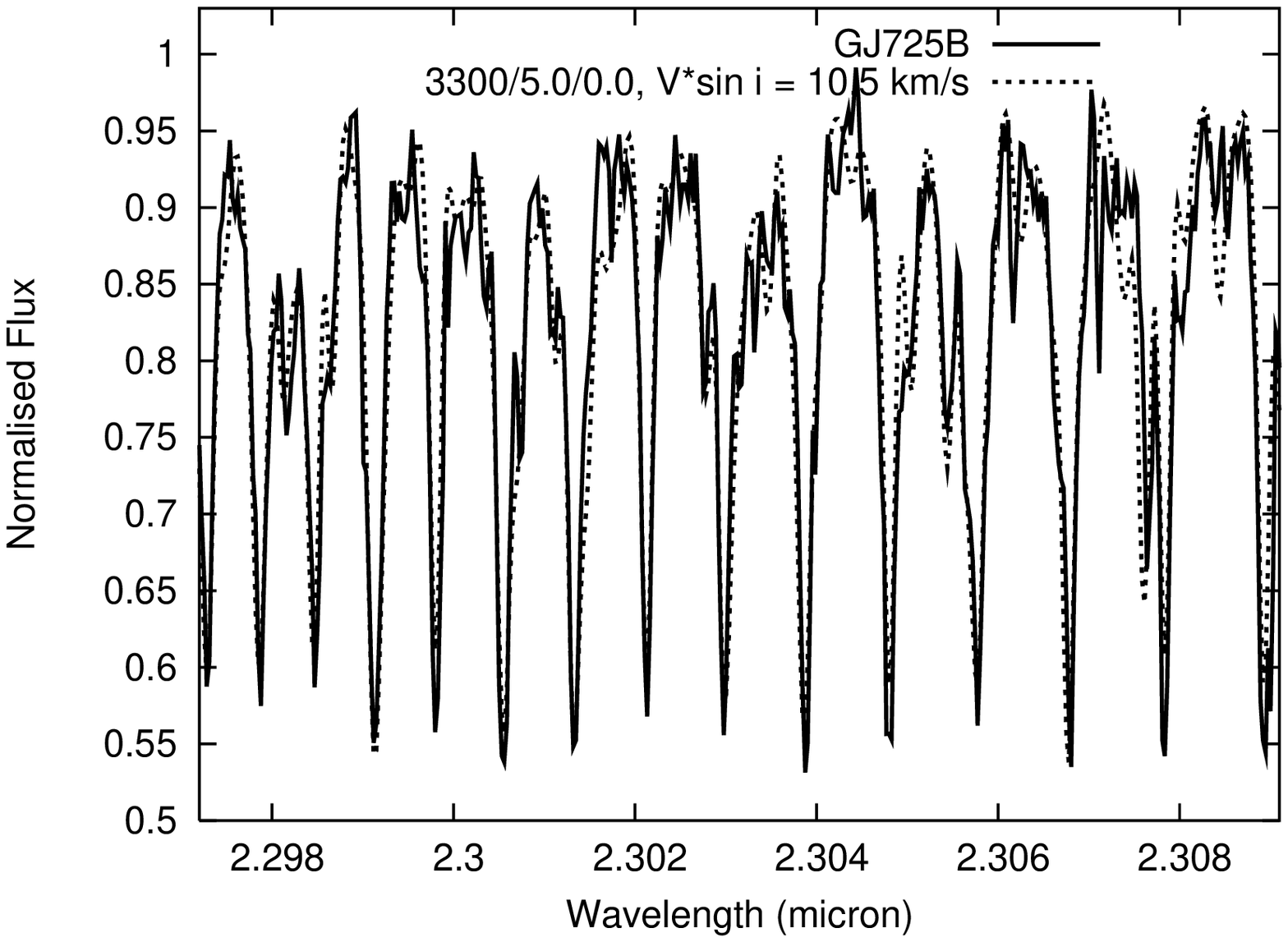}
\includegraphics [width=53mm,angle=0]{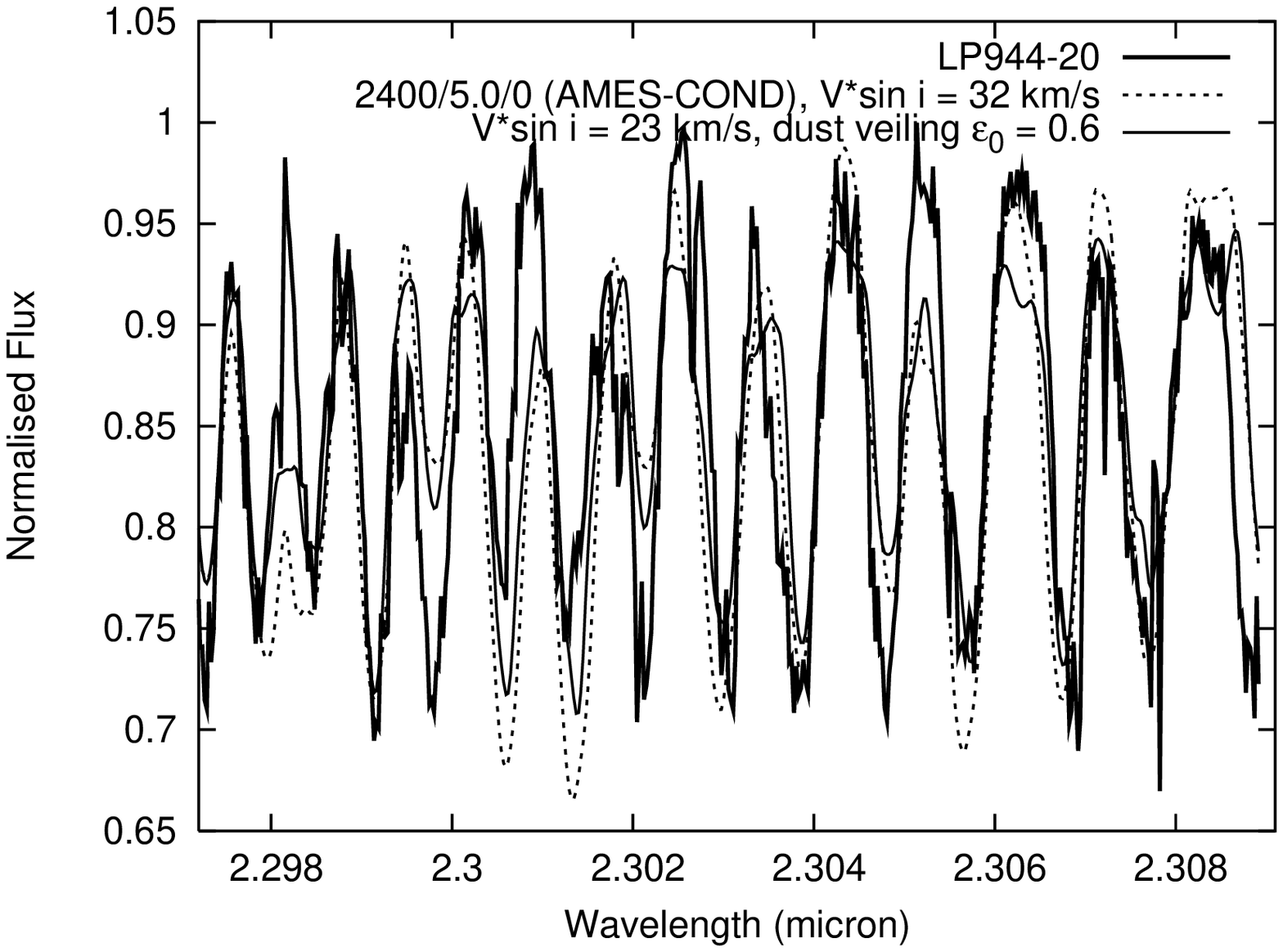}
\includegraphics [width=53mm,angle=0]{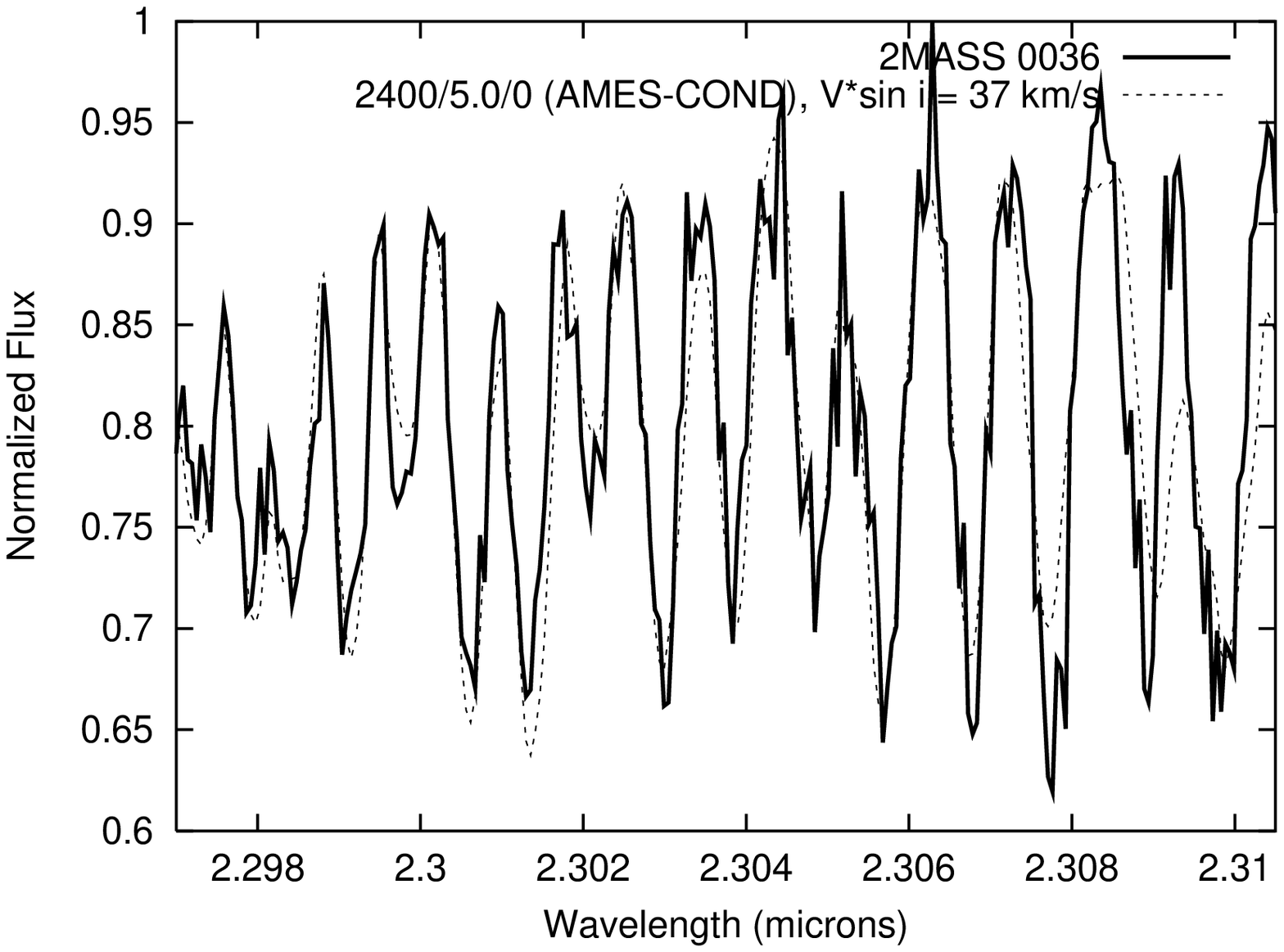}
\end{center}
\caption{The observational data are overlaid with the best fit `unconstrained' 
synthetic spectra except for LP944-20 and 2MASS0036 for which representative 
models are shown.}
\label{fits}
\end{figure*}

\section{The models}
Model atmospheres from the Phoenix code were used for this work. In particular 
we used the grid known as NextGen (Hauschildt et al. 1999) but also Dusty (Allard 
et al. 2001) and AMES-COND models (Allard et al. 2001). Model temperatures of 1500 
to 3800~K, metallicities of [M/H]$=-2.0$ to $0.0$ and gravities of log $g$ = 4.5 to 
5.5 are considered. These parameters represent the probable extremes for the sample 
based on the literature. We have not tried comparing the system with models computed 
with non-solar abundance patterns.

\par
Computations of local thermal equilibrium synthetic spectra were carried out by the 
program WITA6 (Pavlenko 2000). This model assumes LTE, hydrostatic equilibrium for a 
one-dimensional model atmosphere, and no sources or sinks of energy. The equations of 
ionisation-dissociation equilibrium were solved for media consisting of atoms, ions 
and molecules. We took into account $\sim$ 100 components (Pavlenko 2000). The constants 
for equations of chemical balance were taken from Tsuji (1973). It is worth noting that 
the chemical balance in cool dwarf atmospheres is governed by the CO molecule (Pavlenko 
\& Jones 2002).

The Partridge \& Schwenke (1997: PS) line list is used as the primary source of water 
vapour lines, though we also made some comparisons with the preliminary line list of 
Barber \& Tennyson (known as BT1, 2004). The partition functions of \HHO ~were also 
computed on the PS line list following Pavlenko et al. (2004). We recomputed the 
constants of dissociation equilibrium using the \HHO partition function following Vidler 
\& Tennyson (2000) though found no significant differences for test synthetic spectra 
at 3200~K, log $g$~=~5 and 2400~K, log $g$~=~5 (Pavlenko et al, 2004, in preparation). 
For CO, we used the \CC and \CCC ~line lists of Goorvitch (1994). The CO partition 
functions were taken from Gurvitz, Weitz \& Medvedev (1982). The atomic line list was taken from 
VALD (Kupka et al. 1999). The relative importance of the different opacities contributing 
to our synthetic spectra is shown in Fig. \ref{opacities}. 

The profiles of molecular and atomic lines are determined using the Voigt function 
$H(a,v)$, parameters of their natural broadening $C_2$ and van der Waals broadening 
$C_4$ from databases (Kupka et al. 1999) or in their absence computed following Unsold 
(1955). Owing to the low temperatures in cool dwarf atmospheres, and consequent low 
electron densities, Stark broadening may be neglected. Computations for synthetic spectra 
were carried out with a 0.00005 $\mu$m step for microturbulent velocities $v_t$ = 1, 2, 
3 km/s. The sensitivity of the spectral region to changes in model parameters is shown 
in Fig. \ref{sensitivity}. It can be seen that temperature has a relatively larger effect 
on the depth of the CO features than metallicity and gravity. In this regard, a temperature 
change of 200~K is roughly equivalent to a change in metallicity of 1~dex or a change in 
gravity of $\Delta$log~$g$~=~1.

The instrumental broadening was modelled by triangular profiles set to the resolution of 
the observed spectra. To find the best fits to observed spectra we follow the scheme of 
Jones et al. (2002). Namely, for every spectrum we carry out the minimisation of a 3D 
function $S=f(x_{\rm s}, x_{\rm f}, x_{\rm w}) = 1/N \times \sum(1-F_{\rm obs}/F_{\rm synt})^2$, 
where $F_{\rm obs}, F_{\rm synt}$ are observed and computed fluxes, {\em N} is the number of 
points in the observed spectrum to be fitted, and $x_{\rm s}, x_{\rm f}$, and $x_{\rm w}$ are 
relative shifts in wavelength scale, flux normalisation factor, and instrumental $+$ 
rotational broadening, respectively. Rotational broadening was computed following Gray 
(1992). Figure \ref{surface} shows the sensitivity of our fit for GJ752B to the various 
model parameters. Figure \ref{fits} shows the observed and synthetic spectra fit for each 
object.

\begin{figure}
\begin{center}
\includegraphics [width=60mm,angle=90]
{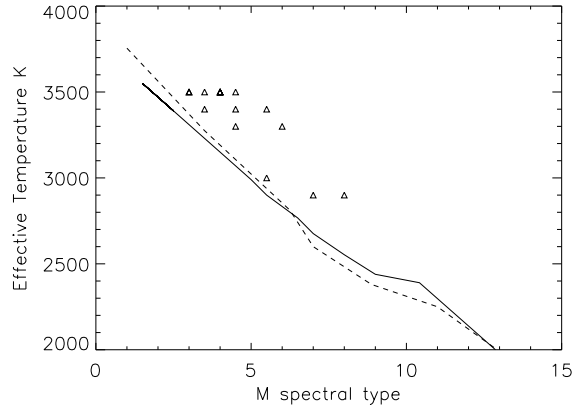}
\end{center}
\caption{The triangles indicate best-fit temperatures for the observed 
sample given synthetic spectra where the metallicities are constrained 
to be solar and gravities constrained to be log~g~=~5.0. The solid line 
shows our adopted `empirical' temperature scale, the dashed line shows 
a recent alternative scale from Golimowski et al. 2004.}
\label{empirical}
\end{figure}

\begin{figure}
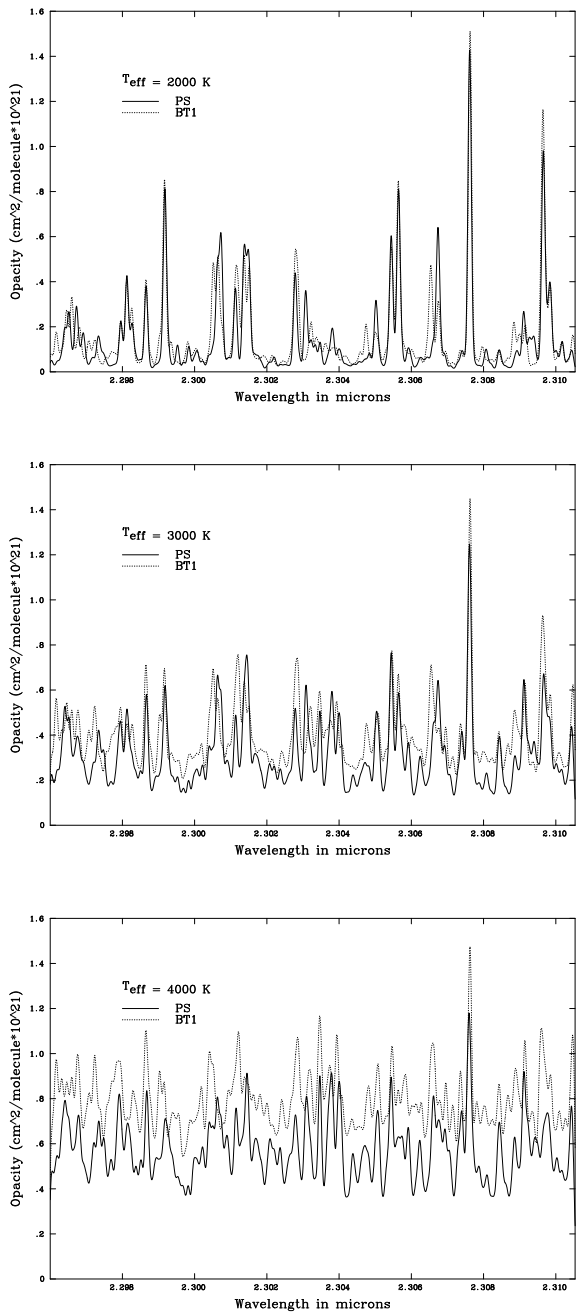

\begin{center}
\includegraphics [width=60mm,angle=-90]
{plot1.eps}
\includegraphics [width=60mm,angle=-90]
{plot2.eps}
\includegraphics [width=60mm,angle=-90]
{plot3.eps}
\end{center}
\caption{Water vapour opacity for different water vapour line lists, the 
solid line shows the Partridge \& Schwenke 1997 (PS) list and the dotted line 
the Barber \& Tennyson 2004 (BT1) list. While across the chosen wavelength 
region they are spectroscopically similar the overall opacity is rather different.}
\label{bt1}
\end{figure}

\section{Spectroscopic analysis}
The effective temperatures given in Table 1 were derived from averaging the temperatures 
derived by Dahn et al. (2002) and Vrba et al. (2004) across each spectral type and 
neglecting very young objects from the determination. For example a much lower effective 
temperature is derived for LP944-21, however, there is strong evidence (e.g. Ribas 2003) 
that the age of this object is $<$ 0.5 Gyr rather than the 3 Gyr assumed by the methodology 
of Dahn et al. (2002) and Vrba et al. (2004). For early spectral types we use the effective 
temperature scales derived from Lane et al. (2001) and Segransan et al. (2003). We note 
these are consistent with detailed studies of GJ630.1A (Viti et al. 2002) and GJ699 (Dawson 
\& de Robertis 2004). From here on we consider the effective temperatures in Table 1 as 
`empirical' though this does assume that the radii of the evolutionary models that were 
used are accurate. Based on the comparisons of Chabrier \& Baraffe (1995) this seems a 
reasonable working assumption.

In Table 2, we present best fit parameters determined using our minimisation technique on 
our observational spectra of sources with spectral types $\le$M8. As suggested by the work 
of Mohanty \& Basri (2003) and Bailer-Jones (2004), it appears that our rotational velocities 
show a general increase toward later spectral types. Although with the modest sample size 
presented here we are not in a position to further advance this area of work. 
Nonetheless, it should be noted that the density of strong CO features makes this an efficient 
region in which to derive rotational velocities and radial velocities. In addition to so-called 
`unconstrained fits' where minimisation takes into account all parameters, we also give two 
cases of `constrained fits'. In the first constrained fit we set [M/H]=0 and fit T$_{eff}$ 
and log $g$, and in the second constrained fit we set T$_{eff}$ equal to the empirical 
T$_{eff}$ and fit log $g$ and [M/H].

Our unconstrained minimisation solutions suggest temperatures higher than would be expected 
from the empirical temperatures of the objects. Relatively high temperatures are also found 
for the [M/H]=0 constrained fit. Our empirical T$_{eff}$ constrained fit indicates a tendency 
to improbably low metallicities for almost all of the sample. We consider this discrepancy to 
arise from the fact that CO bands become weaker for decreasing metallicity or increasing 
temperature in a similar manner. Thus, if the models want to fit a higher T$_{eff}$, but are 
forced to fit a lower one, then they will fit a lower [M/H] to compensate.

In Fig. \ref{empirical} we plot empirical and derived temperature scales against one another. 
It can be seen that for almost all objects our `derived' effective temperatures (triangles) are 
higher than expected for empirical temperatures. Given the good match of synthetic to observed 
spectral features one might expect more similar empirical and synthetic temperature scales. 
Such a discrepancy between empirical and synthetic temperature scale is a long standing result 
which has been arrived at by a number of routes:  from fits to (1) the overall spectral energy 
distribution (e.g., Berriman \& Reid 1987), (2) atomic lines (e.g., Jones et al.  1996), (3) 
water vapour (Jones et al. 2002), (4) carbon monoxide (e.g., Pavlenko \& Jones 2002). However, 
the temperature discrepancy for low-mass stars is usually discussed for stars with T$_{eff}<$3000~K, 
and is generally taken to be an effect arising from the inability of models to accurately treat 
dust formation (e.g. Dahn et al. 2002). We find the temperature discrepancy for apparently 
well-modelled spectra to be prevalent up to at least 3500~K. On the premise that the `empirical' 
temperature scales are closer to reality, the offset suggests that the model structures are too 
hot for a given effective temperature. For GJ752B we have experimented with altering the model 
temperature structure. We find that substantial structural changes are necessary to adjust synthetic 
temperatures by the few hundred K needed to fit empirical ones. Since temperature discrepancies 
of 200 K translate into uncertainties of 1 dex in [M/H] or $\Delta$log $g$=1 it is not possible 
to have confidence in our minimisation values for these properties. The temperature offset for 
the cooler objects such as GJ752B (T$_{eff}$=2550K) might be explained by the presence of relatively 
poorly quantified dust opacities. However, since the offset is also apparent above 3000~K where 
dust will not form  (e.g., Tsuji 2002) it seems likely that at least part of the temperature 
scale problem is not due to dust. Instead we consider that the temperature scale problem may arise 
from a lack of high temperature water opacities.

As expected from our previous high resolution CO work (Viti et al. 2002), our current 
fits of the synthetic to observational spectra are encouraging. This is primarily due to our 
choice of a spectral region dominated by a well understood absorber (CO). However, it is clear 
from Figure \ref{opacities} that water vapour opacity also plays a role in this spectral region. 
While the Partridge \& Schwenke (1997) water vapour line list is clearly excellent at long wavelengths 
(Jones et al. 2002), there are significant discrepancies around 1.6 and 2.2 $\mu$m (Allard, Hauschildt 
\& Schwenke 2000) and most probably around 0.95~$\mu$m as well. 

In Fig. \ref{bt1} we compare the Partridge \& Schwenke (1997) water vapour
line lists with the
preliminary line list of Barber \& Tennyson (2004, in preparation). The
current version of the Barber \& Tennyson
list (known as BT1) is fully converged and complete to J=50 and includes
around 650 million
transitions. The Partridge \& Schwenke list is also complete in respect to
J levels (it reaches J=55) and consists
of around 308 million transitions.   The PS list is cut off at
approximately 28000 cm$^{-1}$ (approximately as there
is not a consistent cut-off level), whereas the BT1 list has a cut-off of
30000 cm$^{-1}$.  Moreover,
even at energies below 28,000 cm$^{-1}$ many lines are missing in the PS
list (probably due to lack
of convergence, resulting in the omission of higher levels). BT1 uses a
newer potential surface (Shirin et al. 2003) and
better describes high temperature water vapour transitions. Figure
\ref {bt1} indicates that at
2000~K the line lists give rather similar results, though by 4000~K BT1
shows approximately a 25\%
increase in opacity. Spectroscopically the line lists are reasonably similar though substantial 
differences can be seen by 4000~K. It is clear that the extra opacity of BT1 arising from higher 
J levels and improved completeness will lead to a significant back-warming effect on the model 
atmosphere moving the photosphere outward and the effective temperature downward. We checked for 
spectral differences between BT1 and PS, however, they are relatively small and make no difference 
to our derived minimisation values. To see the effect of the increased opacity on the atmospheric 
structure of cool dwarfs it will be necessary to compute a new grid of model atmospheres. This will 
begin once the process of checking BT1 is complete, for example, by assigning quantum numbers to 
the transitions of lines in the laboratory emission spectrum of water at 3200~K (Coheur et al. 2004).



\begin{figure}
\begin{center}
\includegraphics [width=80mm,angle=0]
{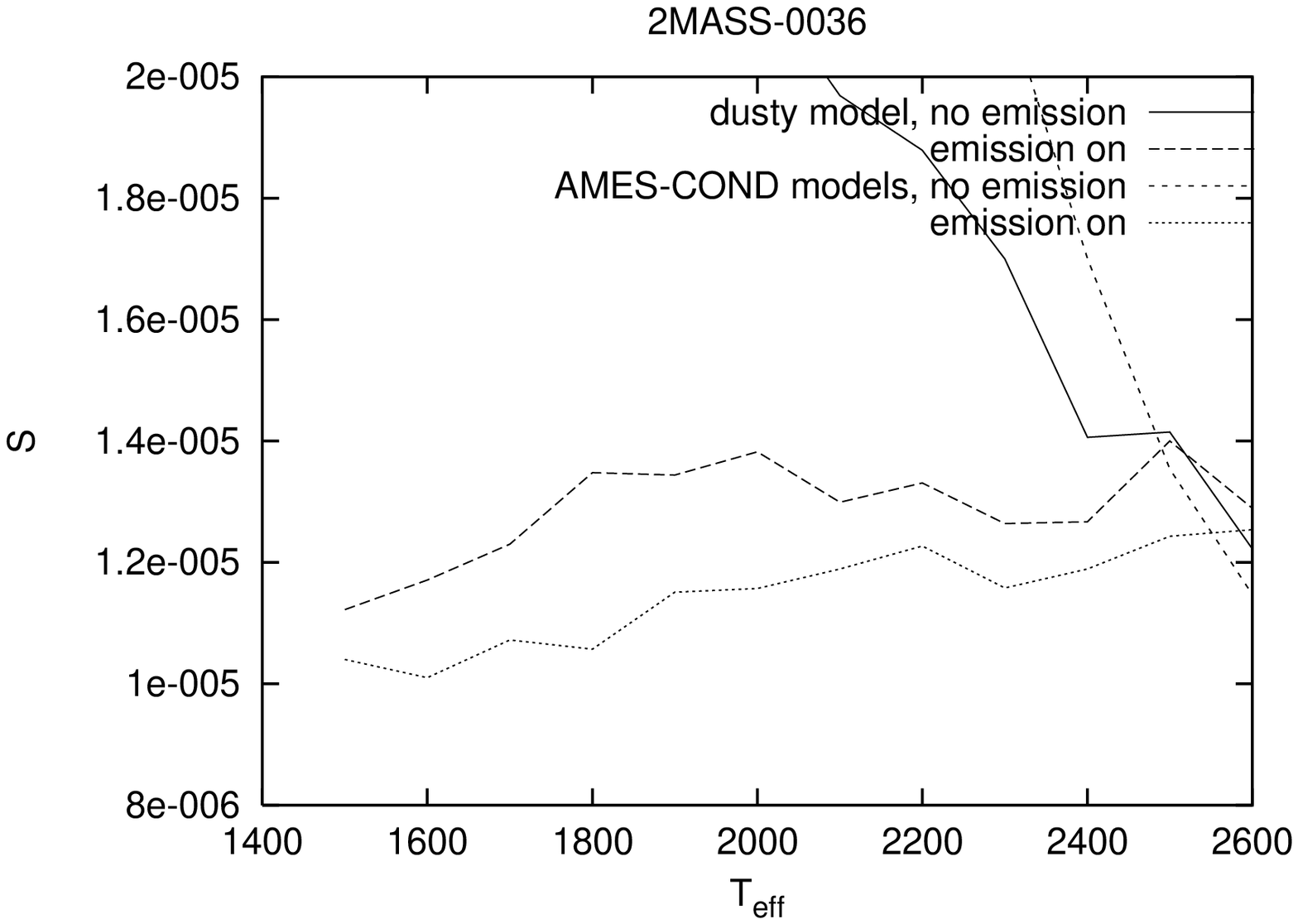}
\includegraphics [width=80mm,angle=0]
{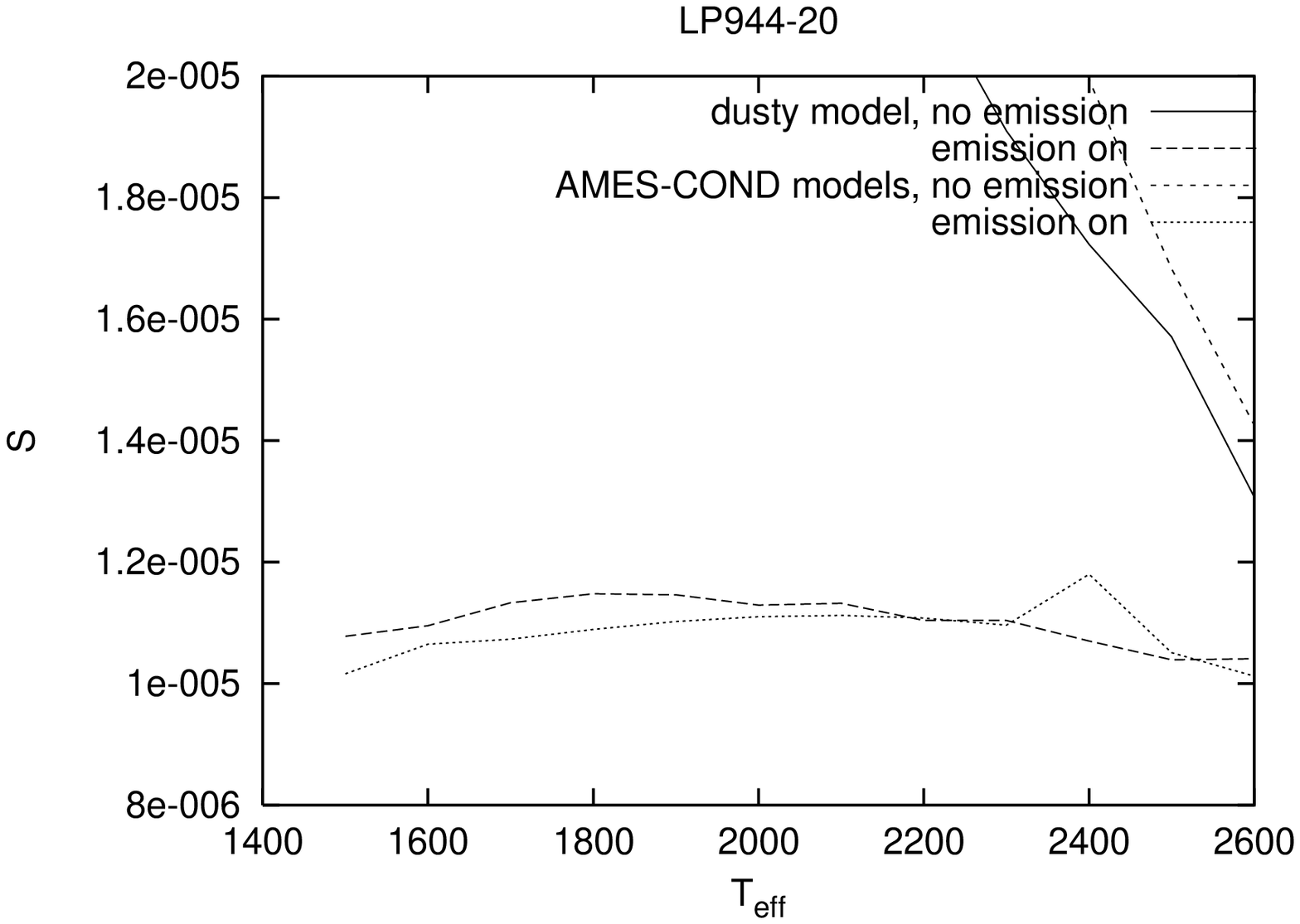}
\end{center}
\caption{The minimisation values for the L dwarfs 2MASS-0036 and LP944 deduced using dusty 
and non-dusty model atmospheres with and without emission.Minimisation values have been 
derived across the relevant model grid of temperatures though log g has been fixed at 5.0 
and [M/H] has been fixed at the solar value.}
\label{ldwarfs}
\end{figure}

\subsection{Below 2500~K -- LP944 and 2MASS0036}\label{coolest}

Below 2500~K we do not easily find a satisfactory minimisation solution for the M9 
dwarf LP944-20 and the L4 dwarf 2M0036, using any of the grids of model atmospheres 
(NextGen, AMES-Cond and Dusty). The results of our experiments are shown in Figure 
\ref{ldwarfs}. The dusty and non-dusty models (solid and short dashed lines respectively) 
both suggest best fit temperatures higher than the edge of the model grid ($>$~2600~K). 
While such temperatures might possibly be plausible for LP944-20, the `empirical' 
temperature for 2MASS0036 is a mere 1900~K. While incomplete water vapour may be 
crucial at higher temperature, the relative similarity of the Partridge \& Schwenke 
and Barber \& Tennyson line lists around 2000~K means that water vapour opacity is 
unlikely to be the source of this discrepancy. The discrepancy is probably more likely 
to arise from an inappropriate treatment of dust opacity. While a more sophisticated 
dust model is certainly appropriate (e.g. Tsuji, Nakajima \& Yanagisawa 2004) we also 
experiment with another possibility. Following the methodology of Pavlenko et al. (2004) 
we imagine that the 2.3 micron region is veiled by additional grey continuum absorption. 
The long-dashed and dotted lines in Figure \ref{ldwarfs} indicate that veiling serves to 
improve the minimisation values obtained and appealingly decreases the sensitivity of the 
model fit. While we have not considered whether the atmospheric conditions are conducive 
to dust emission it is interesting that such veiling could result in spectra where CO 
bands do not change in strength appreciably throughout the L spectral class (Reid et al. 
2001, Geballe et al. 2002, Nakajima, Tsuji \& Yanagisawa 2004). 
Overall, the flat nature of the minimisation values for ``emission on''
models and minimistations beyond the model grid for ``emission off''
models suggest that we are not in a position to constrain effective
temperature for the coolest targets in our sample.

In order to resolve these issues the first priority must be to 
incorporate a new water vapour line list in model atmosphere
calculations. From an observational viewpoint it is important to 
obtain high resolution spectra of appropriate 
molecular and atomic features at widely separated wavelengths.
Such observations should enable us to distinguish the importance
of dust at different wavelengths.
Dust absorption affects
not only the synthetic spectra but the structure of the atmosphere. 
The problem needs to be solved with a  
self-consistent approach which includes among other things: depletion of
molecular species into dust particles, the structure of dust clouds
and a reliable size and composition distribution. 
Consideration of these issues is planned for future models and papers.

\section{Conclusions}
Based on a comparison of high resolution synthetic and observed spectra in the 
near infrared  region, we derive rotational velocities, temperatures, metallicities 
and gravities for a sample of well studied objects. While our spectra are well 
modelled and dominated by CO absorption bands we find temperatures that are higher 
than those found by more empirical methods from 2500-3500~K. The discrepancy at 
higher temperatures is particularly interesting, and we consider it to be indicative 
of missing opacity, probably due to hot water vapour transitions not currently 
included in model atmospheres. Below T$_{eff}$=2500K, the additional complication 
of dust formation is expected to be another factor that requires accurate modelling 
if we are to derive accurate parameters from spectral fits.\\

\section{Acknowledgments}
We thank the staff at the United Kingdom Infrared Telescope for
assistance with the observations, in particular Sandy Leggett.  
We are grateful to PPARC and the Royal Society for travel funding. SV thanks PPARC for an 
advanced fellowship. We are grateful to an anonymous referre for their insightful comments which 
improved this manuscript.


\begin{thebibliography}{99}
\bibitem{}Allard F., Hauschildt P.H., Schwenke D, 2000, ApJ, 540, 1005
\bibitem{}Allard F., Hauschildt P.H., Alexander D.R., Tamanai A., Schweitzer A., 2001, ApJ, 556, 357 
\bibitem{}Bailer-Jones C.A.L., 2004, A\&A, 419, 703
\bibitem{}Berriman G., Reid N., 1987, 227, 315
\bibitem{}Carbon D. F., Milkey R. W., Heasley J. N., 1976, ApJ, 207, 253
\bibitem{}Chabrier G., Baraffe I., 1995, ApJ, 451, 29
\bibitem{}Coheur et al., 2005, J. Chem Phys, in press
\bibitem{}Cushing M. C., Rayner J. R., Davis S. P., Vacca W. D., 2003,
ApJ, 582, 1066 
\bibitem{}Dahn et al. 2002, AJ, 1124, 1170
\bibitem{}Dawson P.C., de Robertis M.M., 2004, AJ, 127, 2909 
\bibitem{}Geballe T.R. et al., 2002, ApJ, 564, 466
\bibitem{}Gizis J.E., 2002, ApJ, 575, 484
\bibitem{}Golimowski, D. A, et al. 2004, astroph/2475
\bibitem{}Goorvitch D., 1994, ApJS, 95, 535
\bibitem{}Gray D.F., 1992, The observation and analysis of stellar photospheres, 2nd edn, Cambridge University Press
\bibitem{}Gurvitz L.V., Weitz I.V., Medvedev V.A., 1982, Thermodynamic properties of individual substances, Moscow Science
\bibitem{} Hauschildt P. H., Allard F., Baron E., 1999, ApJ, 512, 377
\bibitem{}Jones  H. R. A., Longmore A.. J., Allard F., Hauschildt P. H., 1996,
NRAS, 280,77
\bibitem{}Jones H.R.A., Pavlenko Y., Viti S., Tennyson J., 2002, MNRAS, 331, 871
\bibitem{}Kirkpatrick J.D., Henry T., McCarthy D.W., 1991, ApJS, 77, 417 
\bibitem[1999]{_VALD2_}
Kupka, F., Piskunov, N., Ryabchikova, T. A.,
                     Stempels, H. C., Weiss, W. W. 1999,
 A\&AS, 138, 119.
\bibitem{}Lane B.F., Zapatero M.R., Britton M.C., Martin E.L., Kulkarni S.R., 2001, ApJ, 560, 390
\bibitem{}Leggett S. K., 1992, ApJS, 82, 351
\bibitem{}Leggett S. K., Allard F., Hauschildt P H., 1998, ApJ, 509, 836L
\bibitem{}Lyubchik Y, Jones H.R.A., Pavlenko Y. V., Viti S., Pickering J. C., 
Blackwell-Whitehead R., 2004, A\&A, 416, 655        
\bibitem{}McGovern M.R., Kirkpatrick J.D., McLean I.S., Burgasser A.J.,
Prato L., Lowrance P.J., 2004, ApJ, 600, 1020 
\bibitem{}McLean I.S., McGovern M.R., Burgasser A.J., Kirkpatrick J.D., 
Prato L., Kim S.S., 2003, ApJ, 596, 561
\bibitem{}Mohanty S., Basri G., 2003, ApJ, 583, 451 
\bibitem{}Nakajima T., Tsuji T., Yanagisawa K., 2004, ApJ, 607, 499
\bibitem{}Partridge H., Schwenke D.J., 1997, J. Chem. Phys., 106, 4618
\bibitem{}Pavlenko Y., 2000, Astron. Rep., 44, 219
\bibitem{}Pavlenko Y., Geballe T.R., Evans A., Smalley B., Eyres S.P.S., 
Tyne V.H., Yakovina L.A., 2004, A\&A, 417, L39
\bibitem{}Pavlenko Y. \& Jones H.R.A., 2002, A\&A, 396, 967
\bibitem{}Reid I.N., Burgasser A.J., Cruz K.L., Kirkpatrick J.D., 
Gizis J.E., 2001, ApJ, 121, 1710
\bibitem{}Ribas I., 2003, A\&A, 400, 279
\bibitem{}Segransan D., Kervella P., Forveille T., Queloz D.,
2003, A\&A, 397, 5
\bibitem{}Shirin S.V., Polyansky O.L., Zobov N.F., Barletta P., Tennyson J.,
2003, J. Chem. Phys. 118, 2124
\bibitem{}Tsuji T., 1973, 23, 411
\bibitem{}Tsuji T., 2002, ApJ, 575, 264 
\bibitem{}Tsuji T., Nakajima T., Yanagisawa K., 2004, ApJ, 607, 511
\bibitem[1995]{_Unsold1955_}
Unsold, A., 1955  Physik der Sternatmospheren, 2nd ed.
Springer. Berlin.
\bibitem{}Vidler M., Tennyson J., 2000, J. Chem. Phys., 113, 9766
\bibitem{}Viti S., Jones H. R. A., Maxted P., Tennyson J., 2002, MNRAS, 329, 290
\bibitem{}Vrba F.J. et al. 2004, astroph/2272
\end{thebibliography}
\end{document}